\RequirePackage{fix-cm}
\documentclass{svj3modif}                    
\usepackage{graphicx}
 \usepackage{mathptmx}      
 \usepackage{amssymb} 
\newcommand{\sun}  {\odot}
\newcommand{\kms}   {km~s$^{-1}$}

\newcommand{\la}{\lesssim}

\journalname{The Astronomy and Astrophysics Review }

\begin{document}

\title{Radio Jets from Young Stellar Objects}
\subtitle{}


\author{Guillem Anglada  \and Luis F. Rodr\'\i guez 
\and Carlos~Carrasco-Gonz\'alez}


\institute{G. Anglada \at
              Instituto de Astrof\'\i sica de Andaluc\'\i a, CSIC, 
              Glorieta de la Astronom\'\i a s/n, E-18008 Granada, Spain  \\
              \email{guillem@iaa.es}           
           \and
           L. F. Rodr\'\i guez \at
              Instituto de Radioastronom\'\i a y Astrof\'\i sica, UNAM, 
              Apdo. Postal 3-72, 58089 Morelia, Michoac\'an, M\'exico \\
              \email{l.rodriguez@irya.unam.mx} 
            \and 
            C. Carrasco-Gonz\'alez \at 
              Instituto de Radioastronom\'\i a y Astrof\'\i sica, UNAM, 
              Apdo. Postal 3-72, 58089 Morelia, Michoac\'an, M\'exico \\
              \email{c.carrasco@irya.unam.mx}
}

\date{Received: 19 September 2017 / Accepted: 15 March 2018 \\ \\
DOI: 10.1007/s00159-018-0107-z} 



\maketitle

\begin{abstract}

Jets and outflows are ubiquitous in the process of formation of stars 
since outflow is intimately associated with accretion. Free-free 
(thermal) radio continuum emission in the centimeter domain is 
associated with these jets. The emission is relatively weak and compact, 
and sensitive radio interferometers of high angular resolution are 
required to detect and study it.

One of the key problems in the study of outflows is to determine how 
they are accelerated and collimated. Observations in the cm range are 
most useful to trace the base of the ionized jets, close to the young 
central object and the inner parts of its accretion disk, where optical 
or near-IR imaging is made difficult by the high extinction present. 
Radio recombination lines in jets (in combination with proper motions) 
should provide their 3D kinematics at very small scale (near their 
origin). Future instruments such as the Square Kilometre Array (SKA) and 
the Next Generation Very Large Array (ngVLA) will be crucial to perform 
this kind of sensitive observations.

Thermal jets are associated with both high and low mass protostars and 
possibly even with objects in the substellar domain. The ionizing 
mechanism of these radio jets appears to be related to shocks in the 
associated outflows, as suggested by the observed correlation between 
the centimeter luminosity and the outflow momentum rate. From this 
correlation and that of the centimeter luminosity with the bolometric 
luminosity of the system it will be possible to discriminate between 
unresolved HII regions and jets, and to infer additional physical 
properties of the embedded objects.

Some jets associated with young stellar objects (YSOs) show indications 
of non-thermal emission (negative spectral indices) in part of their 
lobes. Linearly polarized synchrotron emission has been found in the jet 
of HH 80-81, allowing one to measure the direction and intensity of the 
jet magnetic field, a key ingredient to determine the collimation and 
ejection mechanisms. As only a fraction of the emission is polarized, 
very sensitive observations such as those that will be feasible with the 
interferometers previously mentioned are required to perform studies in 
a large sample of sources.

Jets are present in many kinds of astrophysical scenarios. 
Characterizing radio jets in YSOs, where thermal emission allows one to 
determine their physical conditions in a reliable way, would also be 
useful in understanding acceleration and collimation mechanisms in all 
kinds of astrophysical jets, such as those associated with stellar and 
supermassive black holes and planetary nebulae.


\end{abstract}

\section{Introduction}
\label{intro}

Until around 1980, the process of star formation was believed to be 
dominated by infall motions from an ambient cloud that made the forming 
star at its center grow in mass. Several papers published at that time 
indicated that powerful bipolar ejections of high-velocity molecular 
(Snell et al. 1980; Rodr\'\i guez et al. 1980) and ionized (Herbig \& 
Jones 1981) gas were also present. The star formation paradigm changed 
from one of pure infall to one in which infall and outflow coexisted. As 
a matter of fact, both processes have a symbiotic relation: the rotating 
disk by which the star accretes provides the energy for the outflow, 
while this latter process removes the excess of angular momentum that 
otherwise will impede further accretion.

Early studies of centimeter radio continuum from visible T Tau stars 
made mostly with the Very Large Array (VLA) showed that emission was 
sometimes detected in them (e.g., Cohen et al. 1982). Later studies made 
evident that in these more evolved stars the emission could have a 
thermal (free-free) origin but that most frequently the emission was 
dominated by a nonthermal (gyrosynchrotron) process (Feigelson \& 
Montmerle 1985) produced in the active magnetospheres of the stars. 
These non-thermal radio stars are frequently time-variable (e.g., 
Rivilla et al. 2015) and its compact radio size makes them ideal for the 
determination of accurate parallaxes using Very Long Baseline 
Interferometry (VLBI) observations (e.g., Kounkel et al. 2017).

In contrast, the youngest low-mass stars, the so-called Class 0 and I 
objects (Lada 1991; Andr\'e et al. 1993) frequently exhibit free-free 
emission at weak but detectable levels (Anglada et al.\ 1992). In the 
best studied cases, the sources are resolved angularly at the sub-arcsec 
scale and found to be elongated in the direction of the large-scale 
tracers of the outflow (e.g., Rodr\'\i guez et al. 1990; Anglada 1996), 
indicating that they trace the region, very close to the exciting star, 
where the outflow phenomenon originates. The spectral index at 
centimeter wavelengths $\alpha$ (defined as $S_\nu \propto \nu^\alpha$), 
usually rises slowly with frequency and can be understood with the 
free-free jet models of Reynolds (1986). Given its morphology and 
spectrum, these radio sources are sometimes referred to as ``thermal 
radio jets''. It should be noted that there are a few Class I objects 
where the emission seems to be dominantly gyrosynchrotron (Feigelson et 
al. 1998; Dzib et al. 2013). These cases might be due to a favorable 
geometry (that is, the protostar is seen nearly pole-on or nearly 
edge-on, where the free-free opacity might be reduced; Forbrich et al.  
2007), or to clearing of circumstellar material by tidal forces in a 
tight binary system (Dzib et al.  2010)

In the last years it has become clear that the radio jets are present in 
young stars across the stellar spectrum, from O-type protostars (Garay 
et al. 2003) and possibly to brown dwarfs (Palau et al. 2014), 
suggesting that the disk-jet scenario that explains the formation of 
solar-type stars extends to all stars and even into the sub-stellar 
regime. The observed centimeter radio luminosities (taken to be $S_\nu 
d^2$, with $d$ being the distance) go from $\sim$100 mJy kpc$^2$ for 
massive young stars to $\sim$$3 \times 10^{-3}$ mJy kpc$^2$ for young 
brown dwarfs.

High sensitivity studies of Herbig-Haro systems revealed that a large 
fraction of them showed the presence of central centimeter continuum 
sources (Rodr\'\i guez \& Reipurth 1998). With the extraordinary 
sensitivity of the Jansky VLA (JVLA) and planed future instrumentation 
it is expected that all nearby (a few kpc) young stellar objects (YSOs) 
known to be associated with outflows, both molecular and/or 
optical/infrared, will be detectable as centimeter sources (Anglada et 
al. 2015).

The topic of radio jets from young stars has been reviewed by Anglada 
(1996), Anglada et al. (2015) and Rodr\'\i guez (1997, 2011). The more 
general theme of multifrequency jets and outflows from young stars has 
been extensively reviewed in the literature, with the most recent 
contributions being Frank et al. (2014) and Bally (2016). In this review 
we concentrate on radio results obtained over the last two decades, 
emphasizing possible lines of study feasible with the improved 
capabilities of current and future radio interferometers.

\section{Information from radio jets}

The study of radio jets associated with young stars has several 
astronomical uses. Given the large obscuration present towards the very 
young stars, the detection of the radio jet provides so far the best way 
to obtain their accurate positions. These observations also provide 
information on the direction and collimation of the gas ejected by the 
young system in the last few years and allow a comparison with the gas 
in the molecular outflows and optical/infrared HH jets, that trace the 
ejection over timescales one to two orders of magnitude larger. This 
comparison allows to make evident changes in the ejection direction 
possibly resulting from precession or orbital motions in binary systems 
(Rodr\'\i guez et al. 2010; Masqu\'e et al. 2012).

In some sources it has been possible to establish the proper motions of 
the core of the jet and to confirm its association with the region 
studied (Loinard et al. 2002; Rodr\'\i guez et al. 2003a; Lim \& 
Takakuwa 2006; Carrasco-Gonz\'alez et al. 2008a; Loinard et al. 2010; 
Rodr\'\i guez et al. 2012a, b). In the case of the binary jet systems 
L1551 IRS5 and YLW 15, the determination of its orbital motions (Lim \& 
Takakuwa 2006; Curiel et al. 2004) allows one to confirm that they are 
solar-mass systems and that they are very overluminous with respect to 
the corresponding main sequence luminosity, as expected for objects that 
are accreting strongly.

The free-free radio emission at cm wavelengths has been used for a long 
time to estimate important physical properties of YSO jets (see below). 
Recently, the observation of the radio emission from jets has reached 
much lower frequencies, as in the recent Low Frequency ARray (LOFAR) 
observations at 149 GHz (2 m) of T Tau (Coughlan et al. 2017). These 
observations have revealed the low-frequency turnover of the free-free 
spectrum, a result that has allowed the degeneracy between emission 
measure and electron density to be broken.

\section{Observed properties of radio jets}

\subsection{Properties of currently known angularly resolved radio jets}

\begin{table}[h!]
{\footnotesize
\caption{Properties of Selected Angularly Resolved Radio Jets in YSOs}
\label{tab:1}
\begin{tabular} 
{l@{\extracolsep{0.4em}}c@{\extracolsep{0.1em}}c@{\extracolsep{0.3em}}c@{\extracolsep{0.3em}}c@{\extracolsep{0.1em}}c@{\extracolsep{0.1em}}c@{\extracolsep{0.1em}}c@{\extracolsep{0.1em}}c@{\extracolsep{0.4em}}c@{\extracolsep{0.4em}}
c@{\extracolsep{0.3em}}c@{\extracolsep{0.3em}}c@{\extracolsep{0.4em}}l}
\hline\noalign{\smallskip}
 {Source}
&  {$L_{\rm bol}$}
&  {$M_\star$}
&  {$d$}
&  {$S_\nu$}
&   
& {$\theta_0$}
& {Size}
& {$v_j$}
& {$t_{\rm dyn}$}
& 
& {$\dot M_{\rm ion}$} 
& {$r_0$} 
& {}\\
& {($L_\odot$)}
& {($M_\odot$)}
& {(kpc)}
& {(mJy)}
& {$\alpha$}
& {(deg)}
& {(au)}
& {(km\,s$^{-1}$)}
& {(yr)} 
& {$\epsilon$}
& {($M_\odot$\,yr$^{-1}$)}
& {(au)}
& {Refs.} \\
\noalign{\smallskip}\hline\noalign{\smallskip}
HH 1-2 VLA1 & 20 & $\sim$1 & 0.4 & 1 & 0.3 
& 19& 200& 270 & 2 & 0.7 & 1$\times$$10^{-8}$ & $\leq$11 & 1, 2, 3, 4 \\
NGC 2071-IRS3 & $\sim$500 & 4 & 0.4 & 3 & 0.6
&40& 200& 400$^a$ & 1 & 1.0 & 2$\times$$10^{-7}$  & $\leq$18 & 5, 6, 2, 7 \\
Cep A HW2 & 1$\times$$10^4$ & 15 & 0.7 & 10 & 0.7  
&14& 400& 460 & 2 & 0.9 & 5$\times$$10^{-7}$ & $\leq$60 & 8, 9, 10, 11, 12 \\
HH 80-81 & 2$\times$$10^4$ & 15 & 1.7 & 5 & 0.2
&34& 1500& 1000 & 4 & 0.6 & 1$\times$$10^{-6}$ & $\leq$25 & 13,\,14,\,15,\,16,\,17,\,18\\
IRAS\,16547-4247 &  6$\times$$10^4$ & 20 & 2.9 & 11 & 0.5 
& 25 & 3000 & 900$^a$ & 8 & 0.9 & 8$\times$$10^{-6}$ & $\leq$310 & 19,\, 20,\, 21,\, 22\\
Serpens &  300 & 3 & 0.42 & 2.8 & 0.2 
& $<$34 & 280 & 300 & 2 & 0.6 & 3$\times$$10^{-8}$ & $\leq$9 & 23,\, 24,\, 25,\, 26,\,27\\
AB Aur & 38 & 2.4 & 0.14 & 0.14 & 1.1 & $<$39 & 24 & 300$^a$ & 0.2 & 3.5 & 2$\times$$10^{-8}$ & $\leq$3 & 28,\,29,\,30\\
L1551 IRS5$^b$ & 20 & 0.6 & 0.14 & 0.8 & 0.1 & 36 & 39 & 150$^a$ & 0.6 & 0.6 & 1$\times$$10^{-9}$ & $\leq$1 & 31,\,32,\,33\\
HH 111$^c$ & 25 & 1.3 & 0.4 & 0.8 & $\sim$1 & $<$79 & 110 & 400 & 0.7 & 2.3 & 2$\times$$10^{-7}$ & $\leq$12 & 34,\,35,\,36,\,37,\,38\\
HL Tau & 7 & 1.3 & 0.14 & 0.3 & $\sim$0.3 & 69 & 27 & 230$^a$ & 0.3 & 0.7 & 2$\times$$10^{-9}$ & $\sim$1.5 & 39,\,40,\,41\\
IC 348-SMM2E & 0.1 & 0.03 & 0.24 & 0.02 & $\sim$0.4 & 45$^d$ & $<$100 & $\sim$50$^a$ & $<$2 & 0.8 & 2$\times$$10^{-10}$ & $\leq$1 & 42,\,43,\,44\\
W75N VLA3 & 750 & 6$^d$  & 2.0 & 4.0 & 0.6 & 37  & 420 & 220 & 4.6 & 1.0  & 6$\times$$10^{-7}$ & $\leq$70 & 45,\,46\\
OMC2 VLA11 & 360 &  4$^d$ & 0.42 & 2.2 & 0.3 & 10  & 200 & 100 & 4.6 & 1.0  & 6$\times$$10^{-7}$ & $\leq$70 & 2,\,47,\,48\\
Re50N & 250 & 4$^d$ & 0.42 & 1.1 & 0.7 & 33 & 450 & 400 & 2.7 &  1.2 & 8$\times$$10^{-8}$ & $\leq$13 & 2,\,49,\,50\\
\noalign{\smallskip}\hline
\end{tabular} \\

$^a$Calculated using eq.\ (\ref{eq:vj}).\\
 $^{b}${Binary twin jet system. The values listed are the mean value of 
the two jets.}\\
 $^{c}${Binary jet system. The values listed are for the dominant 
east-west jet.}\\
 $^{d}${Assumed.} \\
 References: {(1) Fischer et al. 2010; (2) Menten et al. 2007; (3) 
Rodr\'{\i}guez et al.\ 1990; (4) Rodr\'{\i}guez et al.\ 2000; (5) Butner 
et al. 1990; (6) Carrasco Gonz\'alez et al. 2012a; (7) Torrelles et al. 
1998; (8) Hughes et al. 1995; (9) Patel et al. 2005; (10) Dzib et al. 
2011; (11) Curiel et al. 2006; (12) Rodr\'{\i}guez et al.\ 1994b; (13) 
Aspin \& Geballe 1992; (14) Fern\'andez-L\'opez et al. 2011; (15) 
Rodr\'{\i}guez et al.\ 1980b; (16) Mart\'{\i} et al. 1995; (17) 
Mart\'{\i} et al. 1993; (18) Carrasco-Gonz\'alez et al. 2012b; (19) 
Zapata et al. 2015b; (20) Garay et al. 2003; (21) Rodr\'\i guez et al. 
2008; (22) Rodr\'\i guez et al. 2005; (23) Dzib et al. 2010; (24) 
Rodr\'\i guez et al. 1989a; (25) Harvey et al. 1984; (26) Curiel et al. 
1993; (27) Rodr\'\i guez-Kamenetsky et al. 2016; (28) DeWarf et al. 
2003; (29) van den Ancker et al. 1997; (30) Rodr\'\i guez et al. 2014a; 
(31) Liseau et al. 2005; (32) Rodr\'\i guez et al. 2003b; (33) Rodr\'\i 
guez et al. 1998; (34) Reipurth 1989; (35) Lee 2010; (36) G\'omez et al. 
2013; (37) Rodr\'\i guez \& Reipurth 1994; (38) Cernicharo \& Reipurth 
1996; (39) Cohen 1983; (40) ALMA Partnership et al. 2015; (41) A. M. 
Lumbreras et al., in prep; (42) Palau et al. 2014; (43) Forbrich et al. 
2015; (44) Rodr\'\i guez et al. 2014b; (45) Persi et al. 2006; (46) 
Carrasco-Gonz\'alez et al. 2010a; (47) Adams et al. 2012; (48) Osorio et 
al. 2017; (49) Chiang et al. 2015; (50) L. F. Rodr\'\i guez et al., in 
prep.}
 }
\end{table}

\begin{figure}[h!]
\begin{center}
\vspace{0.2cm}
\includegraphics[width=1.0\textwidth]{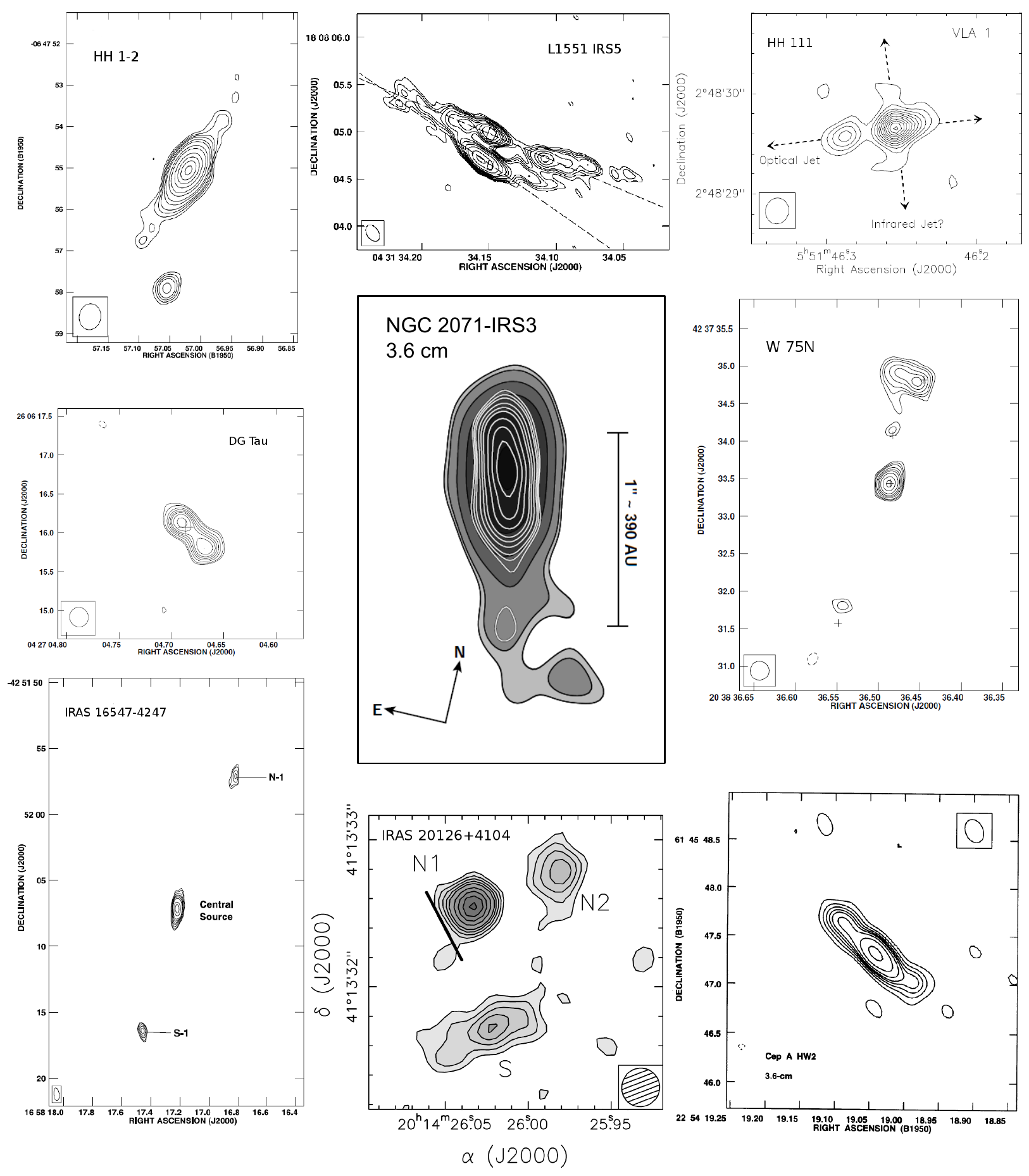}
 \caption{Images of selected radio jets. From left to right, and top to 
bottom: HH 1-2 (Rodr\'\i guez et al. 2000); L1551-IRS5 (Rodr\'\i guez et 
al. 2003b); HH 111 (G\'omez et al. 2013); DG Tau (Rodr\'\i guez et al. 
2012b); NGC2071-IRS3 (Carrasco-Gonz\'alez et al. 2012a); W75N 
(Carrasco-Gonz\'alez et al. 2010a); IRAS 16547$-$4247 (Rodr\'\i guez et 
al. 2005); IRAS 20126+4104 (Hofner et al. 2007); Cep A HW2 (Rodr\'\i 
guez et al. 1994b). Jets from low- and intermediate-mass protostars are 
shown in the top and middle panels, while those from high-mass 
protostars are shown in the bottom panels.
 }
\label{fig:1} 
\end{center}      
\end{figure}

In Table \ref{tab:1} we present the parameters of selected radio jets 
that have been angularly resolved. Several trends can be outlined. The 
spectral index, $\alpha$, is moderately positive, going from values of 
0.1 to $\sim$1, with a median of 0.45. The opening angle of the radio 
jet near its origin, $\theta_0$, is typically in the few tens of 
degrees. In contrast, if we consider the HH objects or knots located 
relatively far from the star, an opening angle an order of magnitude 
smaller is derived. This result has been taken to suggest that there is 
recollimation at scales of tens to hundreds of au. The jet velocity is 
typically in the 100 to 1,000 km s$^{-1}$ range. The ionized mass loss 
rate, $\dot M_{\rm ion}$, is found to be typically an order of magnitude 
smaller than that determined from the large scale molecular outflow, and 
as derived from atomic emission lines, a result that has been taken to 
imply that the radio jet is only partially ionized ($\sim$ 1-10\%; 
Rodr\'\i guez et al. 1990; Hartigan et al. 1994; Bacciotti et al. 1995) 
and that the mass loss rates determined from them are only lower limits. 
In Figure~\ref{fig:1} we show images of several selected radio jets.

\subsection{Proper motions in the jet}

As noted above, comparing observations taken with years of separation it 
has been possible in a few cases to determine the proper motions of the 
core of the jet, whose centroid is believed to coincide within a few 
au's with the young star (e.g., Curiel et al. 2003; Chandler et al. 
2005; Rodr{\'{\i}}guez et al. 2012a, b). These proper motions are 
consistent with those of other stars in the region.

It is also possible to follow the birth and proper motions of new radio 
knots ejected by the star (Mart\'{\i} et al. 1995; Curiel et al. 2006; 
Pech et al. 2010; Carrasco-Gonz\'alez et al. 2010a, 2012a; G\'omez et 
al. 2013; Rodr\'{\i}guez-Kamenetzky et al. 2016; Osorio et al. 2017). 
When radio knots are detected very near the protostar, these are most 
probably due to internal shocks in the jet resulting from changes in the 
velocity of the material with time. These shocks are weak, and the 
emission mechanism seems to be free-free emission from (partially) 
ionized gas. Their proper motions are directly related to the velocity 
of the jet material as it arises from the protostar. The observed proper 
motions of these internal shocks suggest velocities of the jet that go 
from $\sim$100~km~s$^{-1}$ in the low mass stars up to 
$\sim$1000~km~s$^{-1}$ in the most massive objects. So far, in some 
sources there is also evidence of deceleration far from the protostar. 
The radio knots observed by Mart\'\i\ et al. (1995) close to the star in 
the HH 80-81 system move at velocities of $\sim$1000~km~s$^{-1}$ in the 
plane of the sky, while the more distant optical HH objects show 
velocities of order 350~km~s$^{-1}$ (Heathcote et al. 1998; Masqu\'e et 
al. 2015). A similar case has been observed in the triple source in 
Serpens, where knots near the protostar appear to be ejected at very 
high velocities ($\sim$500~km~s$^{-1}$), while radio knots far from the 
protostar move at slower velocities ($\sim$200~km~s$^{-1}$; 
Rodr\'{\i}guez-Kamenetzky et al. 2016). These results suggest that radio 
knots far from the star are then most probably tracing the shocks of the 
jet against the ambient medium. In some cases, when the velocity of the 
jet is high, the emission of these outer knots seems to be of 
synchrotron nature (negative spectral indices), implying that a 
mechanism of particle acceleration can take place at these termination 
shocks. This topic is discussed in more detail below (Section 6).

\subsection{Variability}

Since thermal radio jets are typically detected over scales of $\sim$100 
au and have velocities in the order of 300 km s$^{-1}$, one expects that 
if variations are present they will be detectable on timescales of the 
order of the travel time ($\sim$ a few years) or longer.

The first attempts to detect time variability suggested that modest 
variations, of order 10-20\%, could be present in some sources 
(Mart\'\i\ et al. 1998; Rodr\'\i guez et al. 1999; Rodr\'\i guez et al. 
2000).

However, over time a few examples of more extreme variability were 
detected. The 3.5 cm flux density of the radio source powering the HH 
119 flow in B335 was 0.21$\pm$0.03 mJy in 1990 (Anglada et al. 1992), 
dropping to $\leq$0.08$\pm$0.02 mJy in 1994 (Avila et al. 2001), and 
finally increasing to 0.39$\pm$0.02 mJy in 2001 (Reipurth et al. 2002). 
C. Carrasco-Gonz\'alez et al. (in prep.) report a factor of two increase 
in the 1.0 cm flux density of the southern component of XZ Tau over a 
few months. This increase in radio flux density has been related by 
these authors to an optical/infrared outburst (Krist et al. 2008) 
attributed to the periastron passage in a close binary system.

The radio source associated with DG Tau A presented an important 
increase in its 3.5 cm flux density, that was 0.41$\pm$0.04 mJy in 1994 
to 0.84$\pm$0.05 mJy in 1996 (Rodr\'\i guez et al. 2012b). The radio 
source associated with DG Tau B decreased from a total 3.5 cm flux 
density of 0.56$\pm$0.07 mJy in 1994 to 0.32$\pm$0.05 mJy in 1996 
(Rodr\'\i guez et al. 2012a). In the source IRAS~16293$-$2422A2, Pech et 
al. (2010) observed an increase in the 3.5 cm flux density from 
1.35$\pm$0.08 mJy in 2003 to $>$2.2 mJy en 2009. In these last three 
sources the observed variations were clearly associated with the 
appearance or disappearance of bright radio knots in the systems.

\begin{figure}
\begin{center}
 \includegraphics[width=0.6\textwidth]{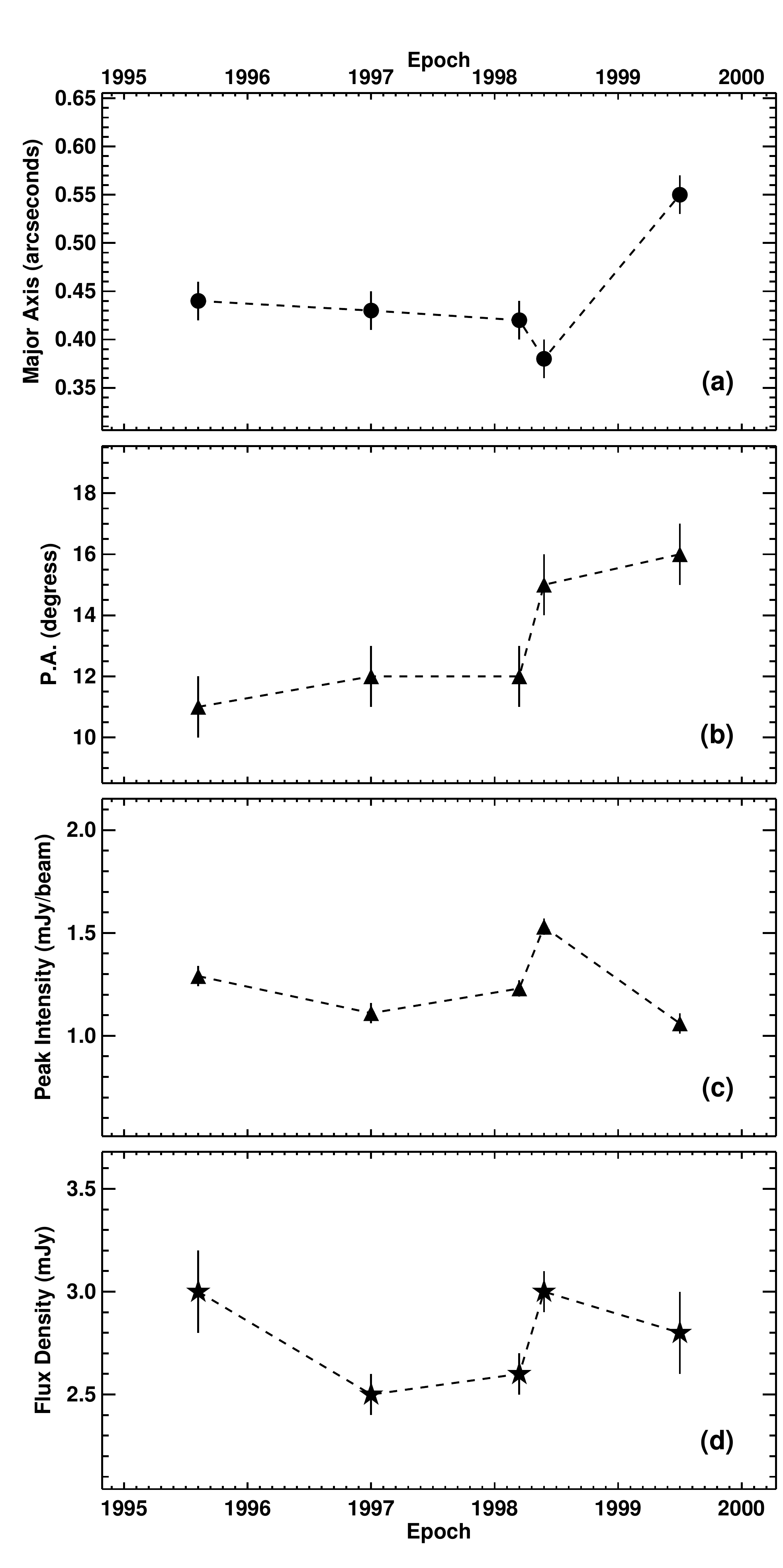}
 \caption{Monitoring of the deconvolved size of the major axis (panel 
a), position angle (panel b), peak intensity (panel c), and flux density 
(panel d) of NGC2071 IRS 3 at 3.6 cm. Image reproduced with permission 
from Carrasco-Gonz\'alez et al. (2012a), copyright by AAS.}
 \label{fig:n2071var}       
\end{center}
\end{figure}

Despite these remarkable cases, most of the radio jets that have been 
monitored show no evidence of variability above the 10-20\% level (e.g., 
Rodr\'\i guez et al. 2008; Loinard et al. 2010; Carrasco-Gonz\'alez et 
al. 2012a, 2015; Rodr\'\i guez et al. 2014a; see 
Fig.~\ref{fig:n2071var}). If ejections are present in these systems, 
they are not as bright as the cases discussed in the previous paragraph.

Given that accretion and outflow are believed to be correlated, the 
variations in the radio continuum emission (tracing the ionized outflow) 
and in the compact infrared and millimeter continuum emissions (tracing 
the accretion to the star) are expected to present some degree of 
temporal correlation. The correlation of the [OI] jet brightness with 
the mid-infrared excess from the inner disk and with the optical excess 
from the hot accretion layer has been considered as evidence that jets 
are powered by accretion (Cabrit 2007 and references therein; see also 
the recent work by Nisini et al. 2018). Also the correlation between the 
radio and bolometric luminosities (see Section 8.1) in very young 
objects supports this hypothesis. In these objects this correlation is 
expected since most of the luminosity comes from accretion, while the 
radio continuum traces the outflow.

However, temporal correlations between the variations in the bolometric 
luminosity and the radio continuum have not been clearly observed. For 
example, the infrared source HOPS 383 in Orion was reported to have a 
large bolometric luminosity increase between 2004 and 2008 (Safron et 
al. 2015), while the radio continuum monitoring at several epochs before 
and after the outburst (from 1998 to 2014) shows no significant 
variation in the flux density (Galv\'an-Madrid et al. 2015). Ellerbroek 
et al. (2014) could not establish a relation between outflow and 
accretion variability in the Herbig Ae/Be star HD 163296, the former 
being measured from proper motions and radial velocities of the jet 
knots, whereas the latter was measured from near-infrared photometric 
and Br$\gamma$ variability. Similarly, Connelley \& Greene (2014) 
monitored a sample of 19 embedded (class I) YSOs with near-IR 
spectroscopy and found that, on average, accretion tracers such as 
Br$\gamma$ are not correlated in time to wind tracers such as the H$_2$ 
and [Fe II] lines. The infrared source LRLL 54361 shows a periodic 
variation in its infrared luminosity, increasing by a factor of 10 over 
roughly one week every 25.34 days. However, the sensitive JVLA 
observations of Forbrich et al. (2015) show no correlation with the 
infrared variations.

\subsection{Outflow rotation}

At the scale of the molecular outflows, several cases have been found 
that show a suggestion of rotation, with a small velocity gradient 
perpendicular to the major axis of the outflow (e.g., Chrysostomou et a. 
2008; Launhardt et al. 2009; Lee et al. 2009; Zapata et al. 2010; Choi 
et al. 2011; Pech et al. 2012; Hara et al. 2013; Chen et al. 2016; 
Bjerkeli et al. 2016). Evidence of outflow rotation has also been found 
in optical/IR microjets from T Tauri stars (e.g., Bacciotti et al. 2002; 
Anderson et al. 2003; Pesenti et al. 2004; Coffey et al. 2004, 2007). 
The observed velocity difference across the minor axis of the molecular 
outflow is typically a few km s$^{-1}$, while in the case of optical/IR 
microjets (that trace the inner, more collimated component) it can reach 
a few tens of km s$^{-1}$. These signatures of jet rotation about its 
symmetry axis are very important because they represent the best way to 
test the hypothesis that jets extract angular momentum from the 
star-disk systems. However, there is considerable debate on the actual 
nature and origin of these gradients (Frank et al. 2014; de Colle et al. 
2016). The presence of precession, asymmetric shocks or multiple sources 
could also produce apparent jet rotation. Also, it is possible that most 
of the angular momentum could be stored in magnetic form, rather than in 
rotation of matter (Coffey et al. 2015). To ensure that the true jet 
rotation is being probed it should be checked that rotation signatures 
are consistent at different positions along the jet, and that the jet 
gradient goes in the same direction as that of the disk. This kind of 
tests have been carried out only in very few objects (see Coffey et al. 
2015) with disparate results. In the case of the rotating outflow 
associated with HH 797 (Pech et al. 2012) a double radio source with 
angular separation of $\sim3''$ is found at its core (Forbrich et al. 
2015), suggesting an explanation in terms of a binary jet system.

It is important that the jet is observed as close as possible to the 
star, where any evidence of angular momentum transfer is still 
preserved, since far from the star the interaction with the environment 
can hide and confuse rotation signatures. Therefore, high angular 
resolution observations of radio recombination lines from radio jets, 
reaching the region closer to the star, could help to solve these 
problems. In principle, it could be possible to observe the jet near its 
launch region and compare the velocity gradients with those observed al 
larger scales. However, these observations are very difficult and can be 
feasible only with new instruments such as the Atacama Large 
Millimeter/submillimeter Array (ALMA) and in the future the Next 
Generation Very Large Array (ngVLA) and the Square Kilometre Array 
(SKA).

\section{Free-free continuum emission from radio jets}

\subsection{Frequency dependences}

Reynolds (1986) modeled the free-free emission from an ionized jet. He 
assumed that the ionized flow begins at an injection radius $r_0$ with a 
circular cross section with initial half-width $w_0$. He adopted 
power-law dependences with radius $r$, so that the half-width of the jet 
is given by

\begin{equation}
w = w_0 \Bigl({{r} \over {r_0}}\Bigr)^\epsilon.
\end{equation}

The case of $\epsilon$ =1 corresponds to a conical (constant opening 
angle) jet. The electron temperature, velocity, density, and ionized 
fraction were taken to vary as $r/r_0$ to the powers $q_T$, $q_v$, 
$q_n$, and $q_x$, respectively. In general. the jets are optically thick 
close to the origin and optically thin at large $r$. The jet axis is 
assumed to have an inclination angle $i$ with respect to the line of 
sight. In consequence, the optical depth along a line of sight through 
the jet follows a power law with index:

 \begin{equation}
q_\tau = \epsilon + 2 q_x + 2 q_n - 1.35 q_T.
 \end{equation}
 Assuming flux continuity
 \begin{equation}
q_\tau = -3 \epsilon + 2 q_x - 2 q_v - 1.35 q_T.
 \end{equation}
 Assuming the case of an isothermal jet with constant velocity and 
ionization fraction ($q_T = q_v = q_x$ = 0), the flux density increases 
with frequency as

\begin{equation}
S_\nu \propto \nu^\alpha,
\end{equation}

\noindent where the spectral index $\alpha$ is given by

\begin{equation}
\alpha = 1.3 - {{0.7} \over {\epsilon}}.
\end{equation}

The angular size of the major axis of the jet decreases with frequency as 

\begin{equation}
\theta_\nu \propto \nu^{-0.7/\epsilon} = \nu^{\alpha - 1.3}.
\end{equation}

The previous discussion is valid for frequencies below the turnover 
frequency $\nu_m$, which is related to the injection radius $r_0$. At 
high enough frequencies, $\nu > \nu_m$, the whole jet becomes optically 
thin and the spectral index becomes $-0.1$.

\subsection{Physical parameters from radio continuum emission}

Following Reynolds (1986) the injection radius and ionized mass loss 
rate are given by:

\begin{eqnarray}\nonumber
\left(\frac{r_0}{\rm au}\right) &=& 
26 \left[\frac{(2-\alpha) (0.1+\alpha)}{1.3-\alpha}\right]^{0.5}
\left[{\left(\frac{S_\nu}{\rm mJy}\right)
\left(\frac{\nu}{\rm 10~GHz}\right)^{-\alpha}}\right]^{0.5}
\left(\frac{\nu_m}{\rm 10~GHz}\right)^{0.5 \alpha - 1} \\
&& \times
\left(\frac{\theta_0 \sin i}{\rm rad}\right)^{-0.5}
\left(\frac{d}{\rm kpc}\right)
\left(\frac{T}{\rm 10^4~K}\right)^{-0.5},
\label{eq:r0}
\end{eqnarray}

\begin{eqnarray}\nonumber
\left(\frac{\dot M_{\rm ion}}{10^{-6}~M_\odot~\rm yr^{-1}}\right)&=&
0.108 \left[\frac{(2-\alpha) (0.1+\alpha)}{1.3-\alpha}\right]^{0.75}
\left[{\left(\frac{S_\nu}{\rm mJy}\right)
\left(\frac{\nu}{\rm 10~GHz}\right)^{-\alpha}}\right]^{0.75} \\
\nonumber
&& \times
\left(\frac{v_{j}}{200~\rm km~s^{-1}}\right)
\left(\frac{\nu_m}{\rm 10~GHz}\right)^{0.75 \alpha - 0.45}
\left(\frac{\theta_0}{\rm rad}\right)^{0.75}
\left({\sin i}\right)^{-0.25} \\
&& \times
\left(\frac{d}{\rm kpc}\right)^{1.5}
\left(\frac{T}{\rm 10^4~K}\right)^{-0.075},
\label{eq:mi}
\end{eqnarray}

\noindent where $\theta_0 = 2 w_0/r_0$ is the injection opening angle of 
the jet, that usually is roughly estimated using

\begin{equation}
\theta_0 = 2~\arctan(\theta_{\rm min}/\theta_{\rm maj}), 
\end{equation}

\noindent where $\theta_{\rm min}$ and $\theta_{\rm maj}$ are the 
deconvolved minor and major axes of the jet. We note that the dimensions 
of the jet are very compact, comparable or smaller than the beam and as 
a consequence the value of $\theta_0$ is uncertain. The opening angle 
determined on larger scales (for example using Herbig-Haro knots along 
the flow away from the jet core) is usually smaller and at present it is 
not known if this is the result of recollimation or of an overestimate 
in the measurement of $\theta_0$.

In the case of a conical ($\alpha$ = 0.6) jet, equations (\ref{eq:r0}) 
and (\ref{eq:mi}) simplify to:

\begin{eqnarray}\nonumber
\left(\frac{r_0}{\rm au}\right) &=& 
31
\left[{\left(\frac{S_\nu}{\rm mJy}\right)
\left(\frac{\nu}{\rm 10~GHz}\right)^{-0.6}}\right]^{0.5}
\left(\frac{\nu_m}{\rm 10~GHz}\right)^{-0.7} \\
&& \times
\left(\frac{\theta_0 \sin i}{\rm rad}\right)^{-0.5}
\left(\frac{d}{\rm kpc}\right)
\left(\frac{T}{\rm 10^4~K}\right)^{-0.5},
\end{eqnarray}

\begin{eqnarray}\nonumber
\left(\frac{\dot M_{\rm ion}}{10^{-6}~M_\odot~\rm yr^{-1}}\right)&=&
0.139
\left[{\left(\frac{S_\nu}{\rm mJy}\right)
\left(\frac{\nu}{\rm 10~GHz}\right)^{-0.6}}\right]^{0.75} \\
\nonumber
&& \times
\left(\frac{v_{j}}{200~\rm km~s^{-1}}\right)
\left(\frac{\theta_0}{\rm rad}\right)^{0.75}
\left({\sin i}\right)^{-0.25} \\
&& \times
\left(\frac{d}{\rm kpc}\right)^{1.5}
\left(\frac{T}{\rm 10^4~K}\right)^{-0.075},
\end{eqnarray}

Usually a value of $T = 10^4$ K is adopted. In a few cases there is 
information on the jet velocity, but in general one has to assume a 
velocity typically in the range from 100 km s$^{-1}$ to 1,000 km 
s$^{-1}$, depending on the mass of the young star. Following Panoglou et 
al. (2012) and assuming a launch radius of about 0.5 au (Estalella et 
al. 2012), the jet velocity can be crudely estimated from

\begin{equation} \label{eq:vj}
\biggl({{v_j} \over {\rm km~s^{-1}}}\biggr) \simeq 140 \biggl({{M_*} 
\over {0.5~M_\odot}}\biggr)^{1/2}.
\end{equation}

Until recently, the turnover frequency has not been determined directly 
from observations and is estimated that it will appear above 40 GHz. 
Only in the case of HL Tau, the detailed multifrequency observations and 
modeling suggest a value of $\sim$1.5 au (A.M. Lumbreras et al., in 
prep.). Determining $\nu_m$ is difficult because at high frequencies 
dust emission from the associated disk starts to become dominant. Future 
observations at high frequencies made with high angular resolution 
should allow to separate the compact free-free emission from the base of 
the jet to that of the more extended dust emission in the disk and give 
more determinations of $\nu_m$ and of the morphology and size of the gap 
between the star and the injection radius of the jet. Anyhow, the 
dependence of $r_0$ on $\nu_m$ is not critical, that is, in order to 
change $r_0$ one order of magnitude, the turnover frequency should 
typically change a factor of $\sim$30. Assuming a lower limit of 
$\sim$10 GHz for the turnover frequency, upper limits of $\sim$10 au 
have been obtained for $r_0$ (Anglada et al. 1998; Beltr\'an et al. 
2001).

In the case of $\dot M_{\rm ion}$ the dependence on $\nu_m$ is almost 
negligible and disappears for the case of a conical jet ($\alpha$ = 
0.6). Using this technique, ionized mass loss rates in the range of 
$10^{-10}~M_\odot$~yr$^{-1}$ (low-mass objects) to 
$10^{-5}~M_\odot$~yr$^{-1}$ (high mass objects) have been determined 
(Rodr\'\i guez et al. 1994b; Beltr\'an et al. 2001; Guzm\'an et al. 
2010, 2012; see Table \ref{tab:1}).

\section{Radio recombination lines from radio jets}

\subsection{LTE formulation}

Following Reynolds (1986), the flux density at frequency $\nu$, $S_\nu$, 
from a jet is given by

\begin{equation}
S_\nu = \int_0^\infty B_\nu(T) (1 - \exp[-\tau_\nu]) d \Omega ,
\end{equation}

\noindent where $B_\nu(T)$ is the source function (taken to be Planck's 
function since we are assuming LTE), $\tau_\nu$ is the optical depth 
along a line of sight through the jet, and $d \Omega$ is the 
differential of solid angle.

Since 
\begin{equation}
d \Omega = {{2 w(r)} \over {d^2}} dy,
\end{equation}

\noindent where $d$ is the distance to the source and $y = r \sin i$ is 
the projected distance in the plane of the sky. Assuming an isothermal 
jet, the continuum emission is given by

\begin{equation}
S_C = B_\nu(T) \int_0^\infty {{2 w(r)} \over {d^2}} (1 - \exp[-\tau_C]) dy,
\end{equation}

\noindent while the line plus continuum emission will be given by

\begin{equation}
S_{L+C} = B_\nu(T) \int_0^\infty {{2 w(r)} \over {d^2}} 
(1 - \exp[-\tau_{L+C}]) dy.
\end{equation}

The line to continuum ratio will be given by

\begin{equation}
{{S_L} \over {S_C}} = {{S_{L+C}} \over {S_C}} -1. 
\end{equation}

Using the power law dependences of the variables and noting that the 
line and the continuum opacities have the same radial dependence, we 
obtain

\begin{equation}
{{S_L} \over {S_C}} = {{\int_0^\infty y^{\epsilon} 
(1 - \exp[-\tau_{L+C}(r_0) y^{\epsilon + 2 q_x + 2 q_n}/ \sin i]) dy}
\over {\int_0^\infty y^{\epsilon} 
(1 - \exp[-\tau_C(r_0) y^{\epsilon + 2 q_x + 2 q_n}/ \sin i]) dy}} - 1.
\end{equation}

Using the definite integral (Gradshteyn \& Ryzhik 1994)

\begin{equation}
\int_0^\infty [1-\exp(-\mu x^p)]~x^{t-1}~ dx =
-{1\over{|p|}}~ \mu^{-{t/p}}~ \Gamma\left({t\over{p}}\right),
\end{equation}

\noindent valid for $0 < t < -p$ for $p < 0$ and with $\Gamma$ being the 
Gamma function.

We then obtain

\begin{equation}
{{S_L} \over {S_C}} = \left[{{\tau_{L+C}(r_0)} \over 
{\tau_C(r_0)}} \right]^{-(\epsilon+1)/(\epsilon + 2 q_x + 2 q_n)} - 1.
\end{equation}

Finally,

\begin{equation}
{{S_L} \over {S_C}} = \left[{{\kappa_L} \over 
{\kappa_C}} + 1 \right]^{-(\epsilon+1)/(\epsilon + 2 q_x + 2 q_n)} - 1.
\end{equation}

\noindent where $\kappa_L$ and $\kappa_C$ are the line and continuum 
absorption coefficients at the frequency of observation, respectively. 
Substituting the LTE ratio of these coefficients (Mezger \& H\"oglund 
1967; Gordon 1969; Quireza et al. 2006), we obtain:


\begin{eqnarray}\nonumber
&&{{S_L} \over {S_C}} = 
\left[0.28 {\left({\nu_L\over{\rm GHz}}\right)^{1.1}}
{\left({{T}\over{\rm 10^4\,K}}\right)^{-1.1}}
{\left(\frac{\Delta v}{\rm km~s^{-1}}\right)^{-1}}{(1 + Y^+)^{-1}}
+1\right]^{-(\epsilon+1)/(\epsilon + 2 q_x + 2 q_n)} -1, \\
&&
\end{eqnarray}

\noindent where $\nu_L$ is the frequency of the line, $\Delta v$ the 
full width at half maximum of the line, and $Y^+$ is the ionized helium 
to ionized hydrogen ratio.

For a standard biconical jet with constant velocity and ionized 
fraction, the equation becomes

\begin{equation}
{{S_L} \over {S_C}} = 
\left[0.28 \biggl({{\nu_L} \over {\rm GHz}} \biggr)^{1.1}
\biggl({{T} \over {\rm 10^4~K}} \biggr)^{-1.1} 
\biggl({{\Delta v} \over {\rm km~s^{-1}}}\biggr)^{-1}
(1 + Y^+)^{-1} + 1 \right]^{2/3} - 1.
\end{equation}

In the centimeter regime, the first term inside the brackets is smaller 
than 1 and using a Taylor expansion the equation can be approximated by

\begin{equation} \label{eq:rrl}
{{S_L} \over {S_C}} = 0.19 \biggl({{\nu_L} \over {\rm GHz}} \biggr)^{1.1}
\biggl({{T} \over {\rm 10^4~K}} \biggr)^{-1.1} 
\biggl({{\Delta v} \over {\rm km~s^{-1}}}\biggr)^{-1}
(1 + Y^+)^{-1}. 
\end{equation}

Are these relatively weak RRLs detectable with the next generation of 
radio interferometers? Let us assume that we attempt to observe the 
H86$\alpha$ line at 10.2 GHz and adopt an electron temperature of $10^4$ 
K and an ionized helium to ionized hydrogen ratio of 0.1. The line width 
expected for these lines is poorly known. If the jet is highly 
collimated, the line width could be a few tens of km s$^{-1}$, similar 
to those observed in HII regions and dominated by the microscopic 
velocity dispersion of the gas. Under these circumstances, we expect to 
see two relatively narrow lines separated by the projected difference in 
radial velocity of the two sides of the jet. However, if the jet is 
highly turbulent and/or has a large opening angle we expect a single 
line with a width of a few hundreds of km s$^{-1}$. Since for a constant 
area under the line it is easier to detect narrow lines (the 
signal-to-noise ratio improves as $\Delta v^{-1/2}$), we will 
conservatively (perhaps pessimistically) assume a line width of 200 km 
s$^{-1}$.

Using equation~(\ref{eq:rrl}) we obtain that the line-to-continuum ratio 
will be $S_L/S_C$ = 0.011. The brightest thermal jets (see Table \ref{tab:1}) have 
continuum flux densities of about 5 mJy, so we expect a peak line flux 
density of 55 $\mu$Jy. These are indeed weak lines. However, the modern 
wide-band receivers allow the simultaneous detection of many 
recombination lines. For example, in the X band (8-12 GHz) there are 11 
$\alpha$ lines (from the H92$\alpha$ at 8.3 GHz to the H82$\alpha$ at 
11.7 GHz). Stacking $N$ different lines will improve the signal-to-noise 
ratio by $N^{1/2}$, so we expect a gain of a factor of 3.3 from this 
averaging.

At present, the Jansky VLA will give, for a frequency of 10.2 GHz, a 
channel width of 100 km s$^{-1}$ and, with an on-source integration time 
of $\sim$45 hours, an rms noise of $\sim$11 $\mu$Jy beam$^{-1}$, 
sufficient to detect the H86$\alpha$ emission at the 5-$\sigma$ level 
from a handful of bright jets. Line stacking will improve this 
signal-to-noise ratio by 3.3. In contrast, the ngVLA is expected to be 
10 times as sensitive as the Jansky VLA, which means that a similar 
sensitivity will be achieved with 1/100 of the time, that is about 0.5 
hours. For an on-source integration time of about 14 hours the ngVLA 
will have an rms noise of $\sim$2 $\mu$Jy beam$^{-1}$ in the line mode 
previously described and will be able to detect the RRLs from the more 
typical 1 mJy continuum jet at the 5-$\sigma$ level. Line stacking will 
increase this detection to a very robust $\sim$16-$\sigma$. This will 
allow to start resolving the jets into at least a few pixels and start 
studying the detailed kinematics of the jet. Then, a program of a few 
hundred hours with the ngVLA will characterize the RRL emission from a 
wide sample of thermal jets. Similar on-source integration times would 
be required with the expected sensitivity of the SKA (see Anglada et al. 
2015).

\subsection{Stark Broadening}

The collisions between the electrons and the atoms produce a line 
broadening, known as Stark or pressure broadening, that depends strongly 
on the electron density of the medium and in particular on the level of 
the transition. For an RRL with principal quantum number $n$, quantum 
number decrement $\Delta n$ in a medium with electron density $N_e$ and 
electron temperature $T_e$, the Stark broadening is given by (Walmsley 
1990):

\begin{equation}
\left(\delta_S \over {\rm km ~s}^{-1} \right) = 
2.72 \left({n \over 42} \right)^{7.5} \Delta n^{-1}
\left({N_e \over 10^7 ~{\rm cm}^{-3}} \right) 
\left({T_e \over 10^4 ~{\rm K}} \right)^{-0.1}.
\end{equation}

Stark broadening has been detected in the case of HII regions (e.g., 
Smirnov et al. 1984; Alexander \& Gulyaev 2016) and slow ionized winds 
(e.g., Guzm\'an et al. 2014). Would it be an important effect in the 
case of RRLs from ionized jets? In contrast with HII regions, where a 
uniform electron density can approximately describe the object, thermal 
jets are characterized by steep gradients in the electron density, that 
is expected to decrease typically as the distance to the star squared 
(see discussion above). On the other hand, the inner parts of a thermal 
jet are optically thick in the continuum and do not contribute to the 
recombination line emission. All the line emission comes from radii 
external to the line of sight where the free-free continuum opacity, 
$\tau_C$, is $\sim$1.

The length of the jets along its major axis is of the order of twice the 
radius at which $\tau_C \simeq 1$. From Table \ref{tab:1} we estimate that at cm 
wavelengths the distance from the star where $\tau_C \simeq 1$ is in the 
range of 100 to 500 au. Assuming a typical opening angle of $30^\circ$ 
we find that the physical depth of the jet, $2 w$, at this position 
ranges from 50 to 250 au. The free-free opacity is given approximately 
by (Mezger \& Henderson 1967):

\begin{equation}
\tau_C \simeq 1.59 \times 10^{-10} 
\left({T_e \over 10^4 ~{\rm K}} \right)^{-1.35}
\left({\nu \over {\rm GHz}}\right)^{-2.1} 
\left({N_e \over {\rm cm}^{-3}}\right)^{2}
\left({2 w \over 100 ~{\rm au}}\right).
\end{equation}

Adopting $T_e$ = $10^4$ K, $\nu$ = 10.2 GHz (the H86$\alpha$ line) and 
$\tau_C$ = 1, we find that the electron density at this position will be 
in the range of $5.7 \times 10^5$~cm$^{-3}$ to $1.3 \times 
10^6$~cm$^{-3}$. The recombination line emission will be coming from 
regions with electron densities equal or smaller than these values. 
Finally, using the equation for the Stark broadening given above, we 
find that the broadening will be in the range of 33 to 167 km s$^{-1}$. 
We then conclude that Stark broadening will be important in the case 
that the line widths are of the order of the microscopic velocity 
dispersion (tens of km s$^{-1}$) but it will not be dominant in the case 
that the line widths have a value similar to that of the jet speed.

\subsection{Non-LTE Radio Recombination Lines}

The treatment presented here assumes that the recombination line 
emission is in LTE, which seems to be a reasonable approach. There are, 
however, cases where the LTE treatment is clearly insufficient and a 
detailed modeling of the line transfer is required. The more clear case 
is that of MWC 349A, a photoevaporating disk located some 1.2 kpc away 
(Gordon 1994). Even when this source is not a jet properly but a 
photoevaporated wind, we discuss it briefly since it is the best case 
known of non-LTE radio recombination lines.

At centimeter wavelengths, the line emission from MWC 349A is consistent 
with LTE conditions (Rodr\'\i guez \& Bastian 1994). However, at the 
millimeter wavelengths, below 3 mm, this source presents strong, 
double-peaked hydrogen recombination lines (Mart\'\i n-Pintado et al. 
1989). These spectra arise from a dense Keplerian-rotating disk, that is 
observed nearly edge-on (Weintroub et al. 2008; B\'aez-Rubio et al. 
2014). The flux density of these features is an order or magnitude 
larger than expected from the LTE assumption and the inference that 
maser emission is at work seems well justified. However, the nature of 
MWC 349A remains controversial, with some groups considering it a young 
object, while others attribute to it an evolved nature (Strelnitski et 
al. 2013).

Extremely broad millimeter recombination lines of a possible maser 
nature have been reported from the high-velocity ionized jet Cep A HW2 
(Jim\'enez-Serra et al. 2011). In this case, the difference between the 
LTE and non-LTE models is less than a factor of two and in some of the 
transitions it is necessary to multiply the non-LTE model intensities by 
a factor of several to agree with the observations. The source Mon 
R2-IRS2 could be a case of weakly amplified radio recombination lines in 
a young massive star. Jim\'enez-Serra et al. (2013) report toward this 
source a double-peaked spectrum. with the peaks separated by about 40 km 
s$^{-1}$. However, the line intensity is only $\sim$50\% larger than 
expected from LTE. Jim\'enez-Serra et al. (2013) propose that the radio 
recombination lines arise from a dense and collimated jet embedded in a 
cylindrical ionized wind, oriented nearly along the line of sight. In 
the case of LkH$\alpha$101, Thum et al. (2013) find the millimeter RRLs 
to be close to LTE and to show non-Gaussian wings that can be used to 
infer the velocity of the wind, in this case 55 km s$^{-1}$. Finally, 
Guzm\'an et al. (2014) detect millimeter recombination lines from the 
high-mass young stellar object G345.4938+01.4677. The hydrogen 
recombination lines exhibit Voigt profiles, which is a strong signature 
of Stark broadening. The continuum and line emissions can be 
successfully reproduced with a simple model of a slow ionized wind in 
LTE.

\section{Non-thermal emission from radio jets}

\begin{figure}
\begin{center}
\includegraphics[width=0.9\textwidth]{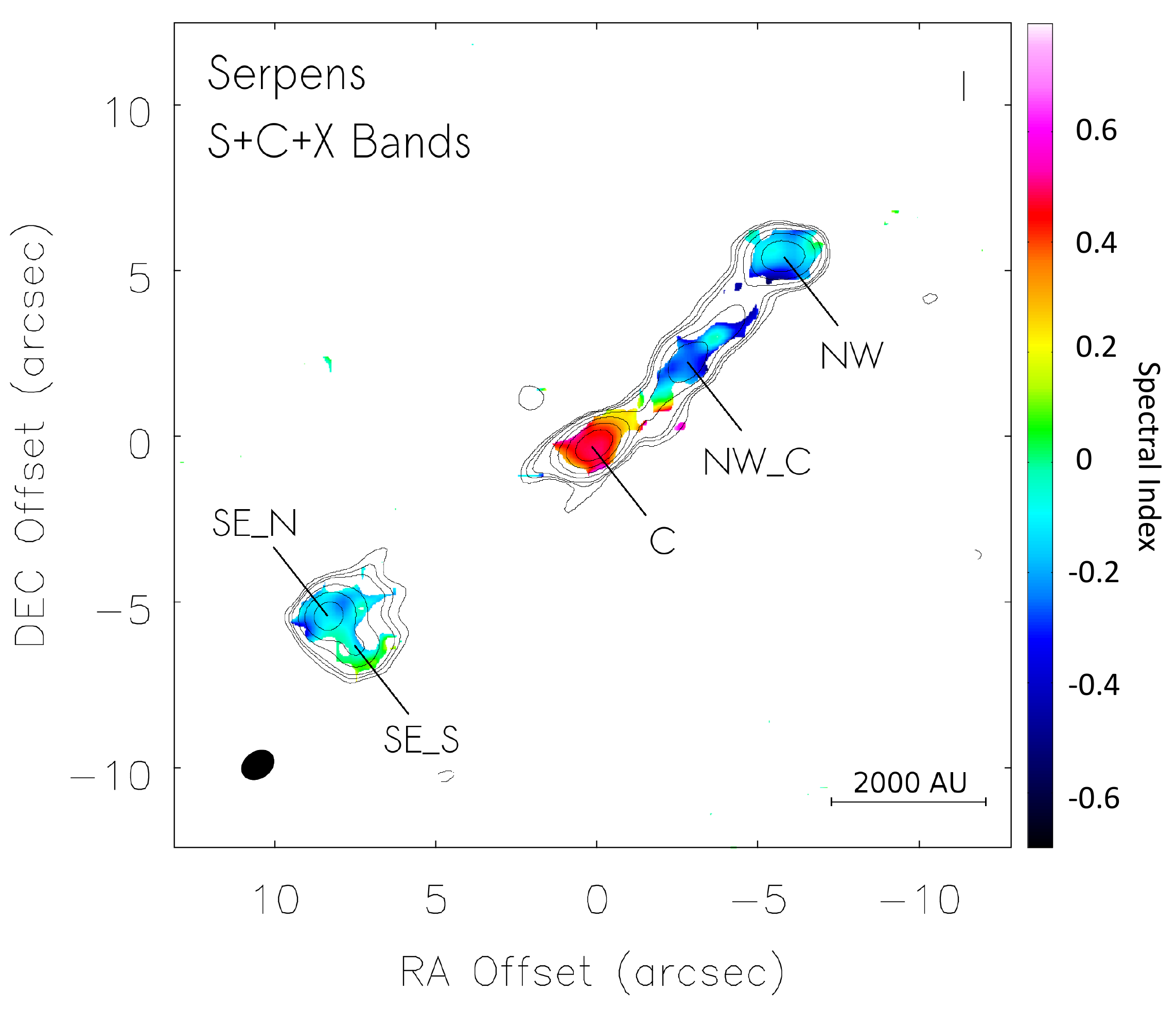}
 \caption{Image of the non-thermal radio jet in Serpens. The radio 
continuum image is shown in contours and was made by combining the data 
from the S, C, and X bands. This image is superposed on the spectral 
index image (color scale). Image reproduced with permission from 
Rodr\'\i guez-Kamenetzky et al.\ (2016), copyright by AAS.
 }
\label{fig:serpens}       
\end{center}
\end{figure}

 As commented above, jets from YSOs have been long studied at radio 
wavelengths through their thermal free-free emission which traces the 
base of the ionized jet, and shows a characteristic positive spectral 
index (the intensity of the emission increases with the frequency, e.g., 
Rodr\'\i guez 1995; Rodr\'\i guez 1996; Anglada 1996). However, in the 
early 1990s, sensitive observations started to reveal that non-thermal 
emission at centimeter wavelengths might be also present in several YSO 
jets (e.g., Rodr\'\i guez et al. 1989a; Rodr\'\i guez et al. 2005; 
Mart\'{\i} et al. 1993; Garay et al 1996; Wilner et al. 1999). This 
non-thermal emission is usually found in relatively strong radio knots, 
away from the core, showing negative spectral indices at centimeter 
wavelengths (see Fig.~\ref{fig:serpens}). Oftentimes, these non-thermal 
radio knots appear in pairs, moving away from the central protostar at 
velocities of several hundred kilometers per second (e.g., the triple 
source in Serpens; Curiel et al. 1993). Because of these 
characteristics, it was proposed that these knots could be tracing 
strong shocks of the jet against dense material in the surrounding 
molecular cloud where the protostar is forming. Their non-thermal nature 
was interpreted as synchrotron emission from a small population of 
relativistic particles that would be accelerated in the ensuing strong 
shocks. This scenario is supported by several theoretical studies, 
showing the feasibility of strong shocks in protostellar jets as 
efficient particle accelerators (e.g., Araudo et al. 2007, Bosch-Ramon 
et al. 2010, Padovani et al. 2015, 2016, C\'ecere et al. 2016).

\begin{figure}
\begin{center}
\includegraphics[width=0.9\textwidth, angle = 0]{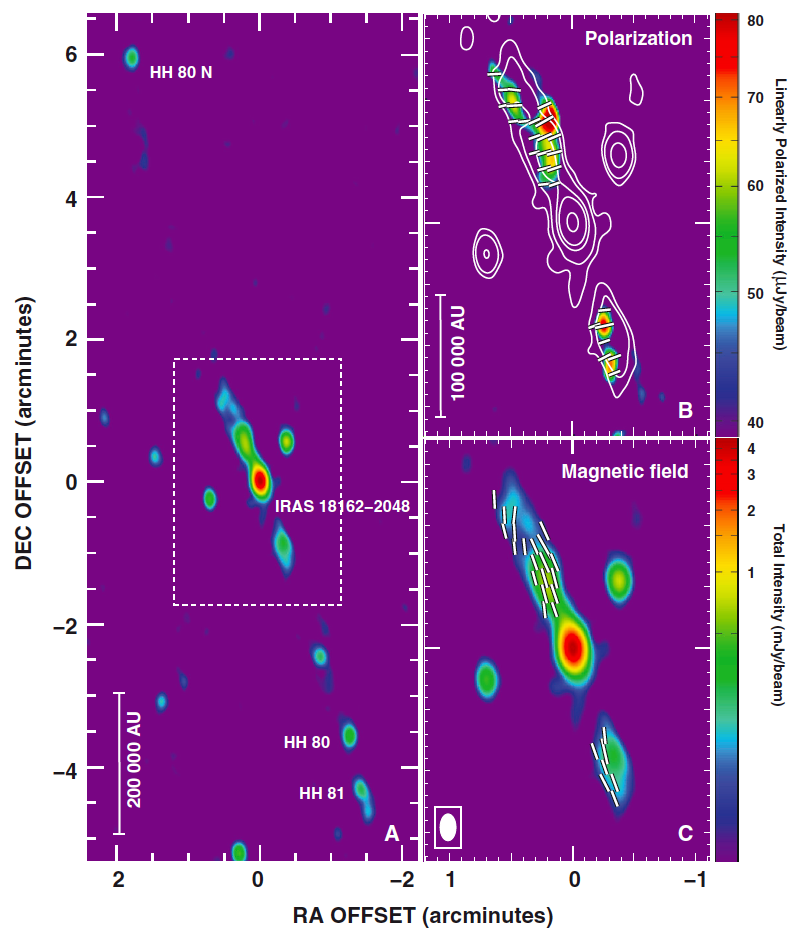}
 \caption{Images of the non-thermal HH 80-81 radio jet. The left panel 
(A) shows in color scale the total intensity at 6 cm extending from the 
central source to the HH objects. The upper right panel (B) shows a 
close-up of the central region with the total intensity in contours and 
the linearly polarized emission in colors. The white bars indicate the 
direction of the polarization. The bottom right panel (C) shows the 
direction of the magnetic field. Image reproduced with permission from 
Carrasco-Gonz\'alez et al.\ (2010b), copyright by AAAS.
 }
\label{fig:hh80}       
\end{center}
\end{figure}

 The possibility of protostellar jets as efficient particle accelerators 
has been confirmed only recently, with the detection and mapping of 
linearly polarized emission from the YSO jet in HH~80-81 
(Carrasco-Gonz\'alez et al. 2010b; see Fig.~\ref{fig:hh80}). This result 
provided for the first time conclusive evidence for the presence of 
synchrotron emission in a jet from a YSO, and then, the presence of 
relativistic particles.

\begin{figure}
\begin{center}
\includegraphics[width=1.0\textwidth, angle = 0]{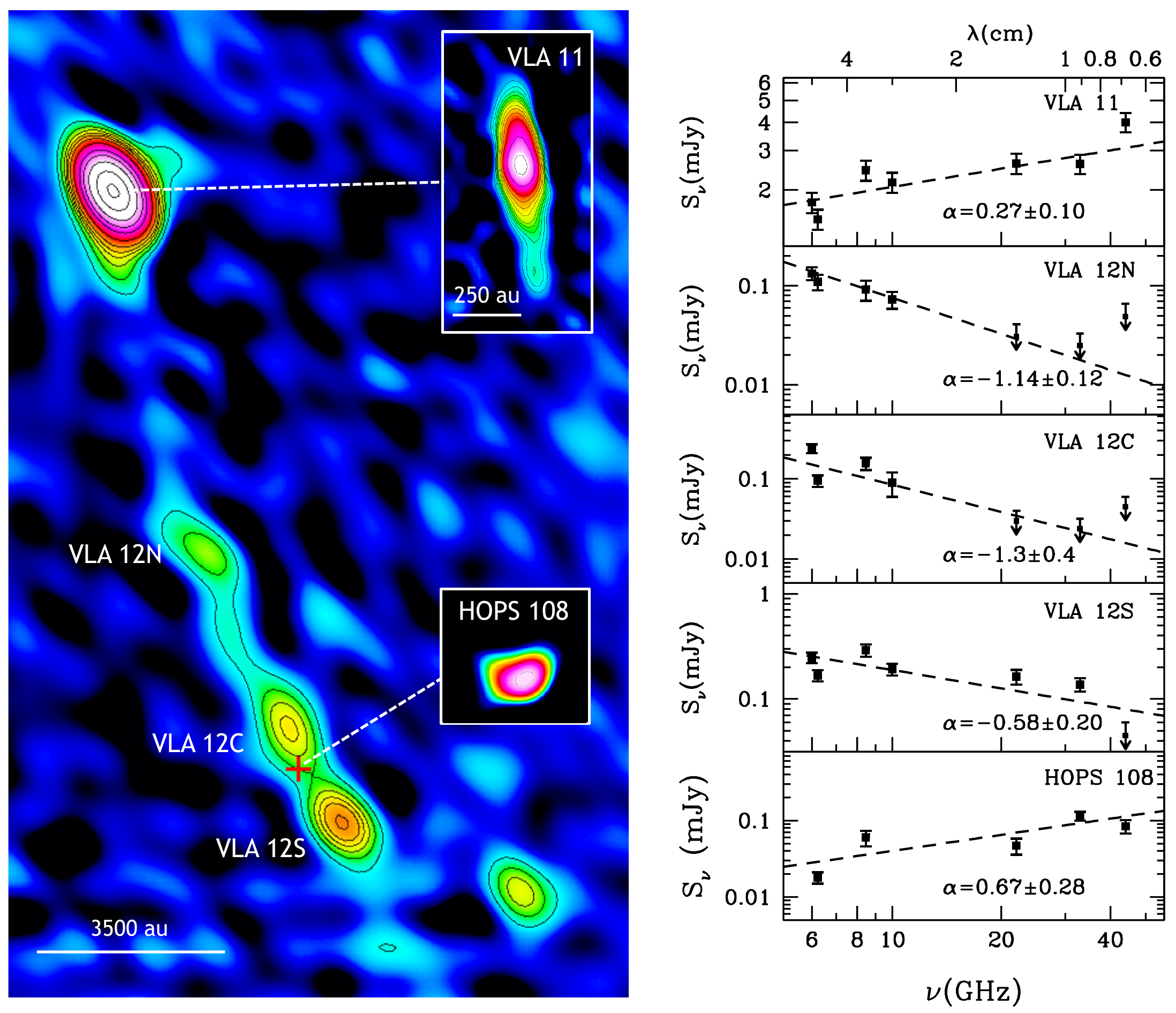}
 \caption{The non-thermal radio jet from the intermediate-mass YSO FIR3 
(VLA 11) in OMC-2. The left panel shows the 3 cm continuum emission 
(angular resolution $\sim 2''$) of the thermal core (VLA 11) and the 
non-thermal lobe (VLA 12N, 12C, 12S) of the FIR3 radio jet. Insets show 
the 5 cm emission at higher angular resolution ($\sim 0.3''$) of the 
thermal core of the FIR3 radio jet and of the Class 0 protostar HOPS 
108. The right panels show the spectra of the radio sources. Image 
adapted from Osorio et al.\ (2017).
 }
\label{fig:hops108}       
\end{center}
\end{figure}

At centimeter wavelengths, where most of the radio jet studies have been 
carried out, the emission of the non-thermal lobes is usually weaker 
than that of the thermal core of the jet. Consequently, most of the 
early detections of non-thermal radio knots were obtained through very 
sensitive observations resulting from unusually long projects, mainly 
carried out at the VLA. After the confirmation in 2010 of synchrotron 
emission in the HH 80-81 radio jet, the improvement in sensitivity of 
radio interferometers has facilitated higher sensitivity observations at 
multiple wavelengths of star forming regions, and new non-thermal 
protostellar jet candidates are emerging (e.g., Purser et al. 2016, 
Osorio et al. 2017, Hunter et al. 2018; Tychoniec et al. 2018; see 
Fig.~\ref{fig:hops108}). It is then expected that the next generation of 
ultra-sensitive radio interferometers (LOFAR, SKA, ngVLA) will produce 
very detailed studies on the nature of non-thermal emission in 
protostellar jets.

\begin{figure}
\begin{center}
\includegraphics[width=0.9\textwidth, angle = 0]{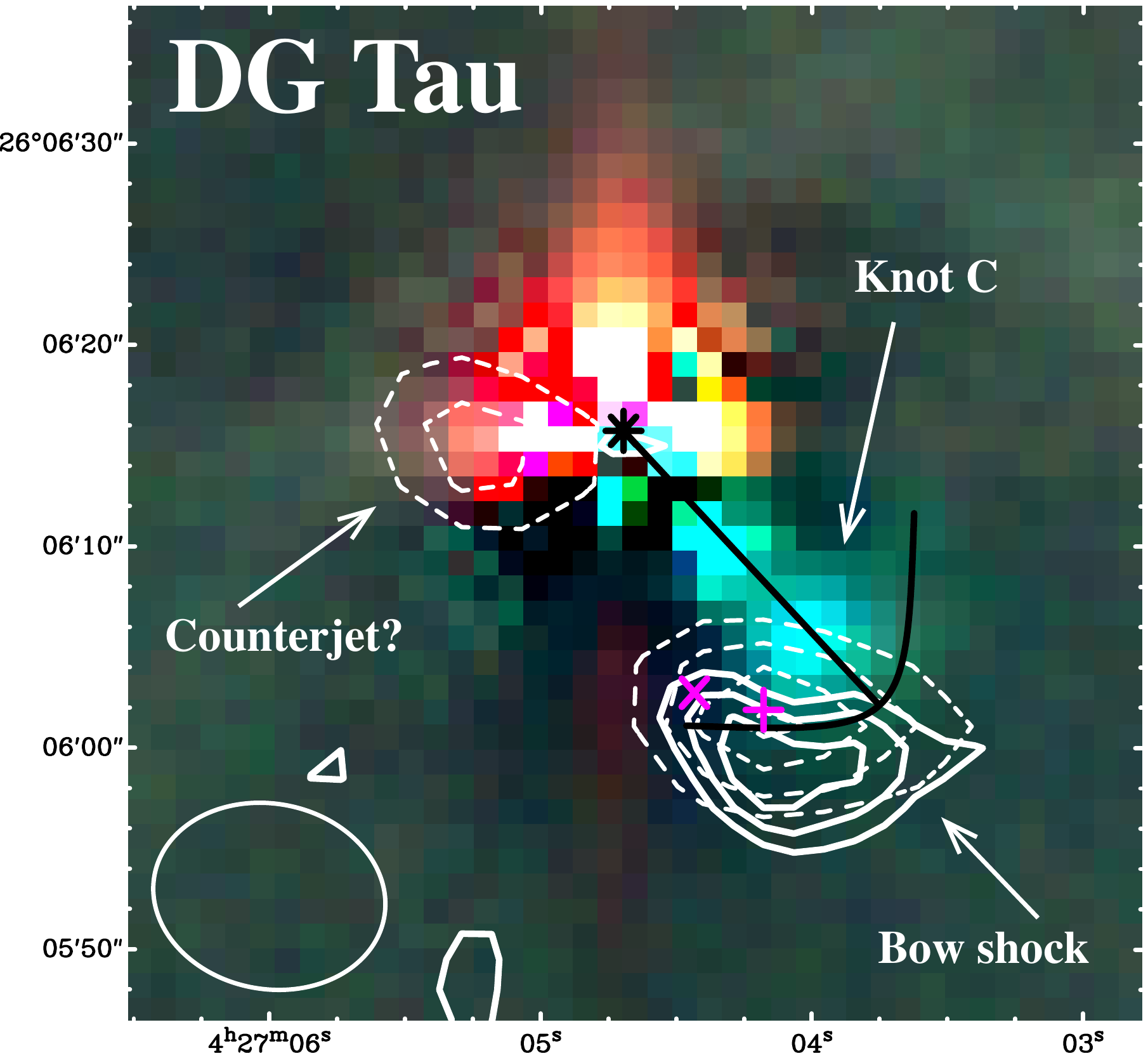}
 \caption{GMRT image at 325 MHz (dashed contours) and 610 MHz (solid 
contours) overlaid on a composite RGB image (I, H$\alpha$ and [SII]) of 
the DG Tau jet. The VLA positions of the bow shock at 5.4 and 8.5 GHz 
are shown as a plus (+) and a cross ($\times$), respectively. The 
optical stellar position is shown as an asterisk ($*$) and the optical 
jet axis and bow shock are shown as solid black lines. Image reproduced 
with permission from Ainsworth et al. (2014), copyright by AAS.
 }
\label{fig:dgtau}       
\end{center}
\end{figure}

 It is worth noting that, even when non-thermal emission is brighter at 
lower frequencies, most of the proposed detections have been performed 
at relatively high frequencies, typically in the 4-10 GHz range. One 
reason is that studies of radio jets have been focused preferentially on 
the detection of thermal emission, which is dominant at higher 
frequencies. Another reason is the lack of sensitivity at low 
frequencies, specially below 1 GHz, in most of the available 
interferometers. However, despite the small number of sensitive 
observations performed at low frequencies, they have produced very 
interesting results. For example, the only non-thermal radio jet 
candidate associated with a low-mass protostar known so far has been 
identified through Giant Metrewave Radio Telescope (GMRT) observations 
of DG Tau in the 300-700 MHz range (Ainsworth et al. 2014; see 
Fig.~\ref{fig:dgtau}). Very recent GMRT observations of HH 80-81 (Vig et 
al. 2018) at 325 and 610 MHz found negative spectral indices steeper 
than previous studies at higher frequencies. This has been interpreted 
as indicating that an important free-free contribution is present at 
high frequencies even in the non-thermal radio knots. We should then 
expect more interesting results from low frequency observations using 
the next generation of low frequency interferometers, such as the Low 
Wavelength Array (LWA), and the aforementioned LOFAR and SKA.

 At present, the best studied cases of particle acceleration in 
protostellar jets are the triple source in Serpens and HH 80-81 
(Rodr\'{\i}guez-Kamenetzky et al. 2016, 2017). These sources were 
observed very recently with the VLA combining high angular resolution 
and very high sensitivity in the 1-10 GHz frequency range. These 
observations resolve both jets at all observed frequencies, allowing to 
study the different emission mechanisms present in the jet in a 
spatially resolved way. In both cases, strong non-thermal emission is 
detected at the termination points of the jets, which is consistent with 
particle acceleration in strong shocks against a very dense ambient 
medium. Additionally, in the case of HH 80-81, non-thermal emission is 
observed at several positions along the very well collimated radio jet, 
suggesting that this object is able to accelerate particles also in 
internal shocks (Rodr\'{\i}guez-Kamenetzky et al. 2017). Overall, these 
studies have found that the necessary conditions to accelerate particles 
in a protostellar jet are high velocities ($\gtrsim$ 500 km~s$^{-1}$) 
and an ambient medium denser than the jet. These conditions are probably 
well satisfied in the case of very young protostellar jets, which are 
still deeply embedded in their parental cloud.

 The discovery of linearly polarized emission in the HH 80-81 radio jet 
by Carrasco-Gonz\'alez et al. (2010b) opened the interesting possibility 
of studying magnetic fields in protostellar jets. Detection of linearly 
polarized emission at several wavelengths would allow one to infer the 
properties of the magnetic field in these jets in a way similar to that 
commonly employed in AGN jets. The magnetic field strength can be 
estimated from the spectral energy distribution at cm wavelengths (e.g., 
Pacholczyk 1970; Beck \& Krause 2005), while the magnetic field 
morphology can be obtained from the properties of the linear 
polarization (polarization angle, polarization degree and Faraday 
rotation). For non-relativistic jets, the apparent magnetic field 
(magnetic field averaged along the line-of-sight) is perpendicular to 
the direction of the linear polarization. Moreover, theoretical models 
of helical magnetic fields predict gradients of the polarization degree 
and Faraday rotation measurements along and across the jet (Lyutikov et 
al. 2005). Thus, by comparing the observational results with theoretical 
models, the 3D morphology of the magnetic field can be inferred.
 
 Mapping the polarization in a set of YSO jets in combination with 
detailed theoretical modeling may lead to a deeper understanding of the 
overall jet phenomenon. However, radio synchrotron emission in YSO jets 
seems to be intrinsically much weaker and difficult to study than in 
relativistic jets (e.g., AGN and microquasar jets). So far, only in the 
case of HH 80-81, one of the brightest and most powerful YSO jets known, 
it has been possible to obtain enough sensitivity to detect and study 
its linear polarization, which is only a fraction of the total continuum 
emission. Subsequent higher sensitivity studies 
(Rodr\'{\i}guez-Kamenetzky et al. 2016, 2017) have not been able to 
detect polarization in YSO jets. One of the reasons of this is that, due 
to the presence of thermal electrons mixed with the relativistic 
particles, we expect strong Faraday rotation. At the moment, high 
sensitivities in radio interferometers are obtained by averaging data in 
large bandwidths. Within these bandwidths, we expect large rotations of 
the polarization angle, resulting in strong depolarization when the 
emission is averaged. Then, it is still necessary to observe using long 
integration times in order to obtain enough sensitivity to detect 
polarization in smaller frequency ranges. Note that using the Faraday 
RM-Synthesis tool (Brentjens \& de Bruyn 2005) to determine the rotation 
measure to account for this effect is also hampered by the poor 
signal-to-noise ratio over narrow frequency ranges in these weak 
sources. At the moment, probably HH 80-81 is the only object bright 
enough to perform a study of polarization at several wavelengths; and 
even in this object, polarization would be detected only after 
observations of several tens of hours long. Therefore, we will probably 
have to wait for new ultra-sensitive interferometers in order to perform 
a polarization study in a large sample of protostellar jets.

\section{Masers as tracers of jets}

\begin{figure}[h!]
\begin{center}
\includegraphics[width=\textwidth]{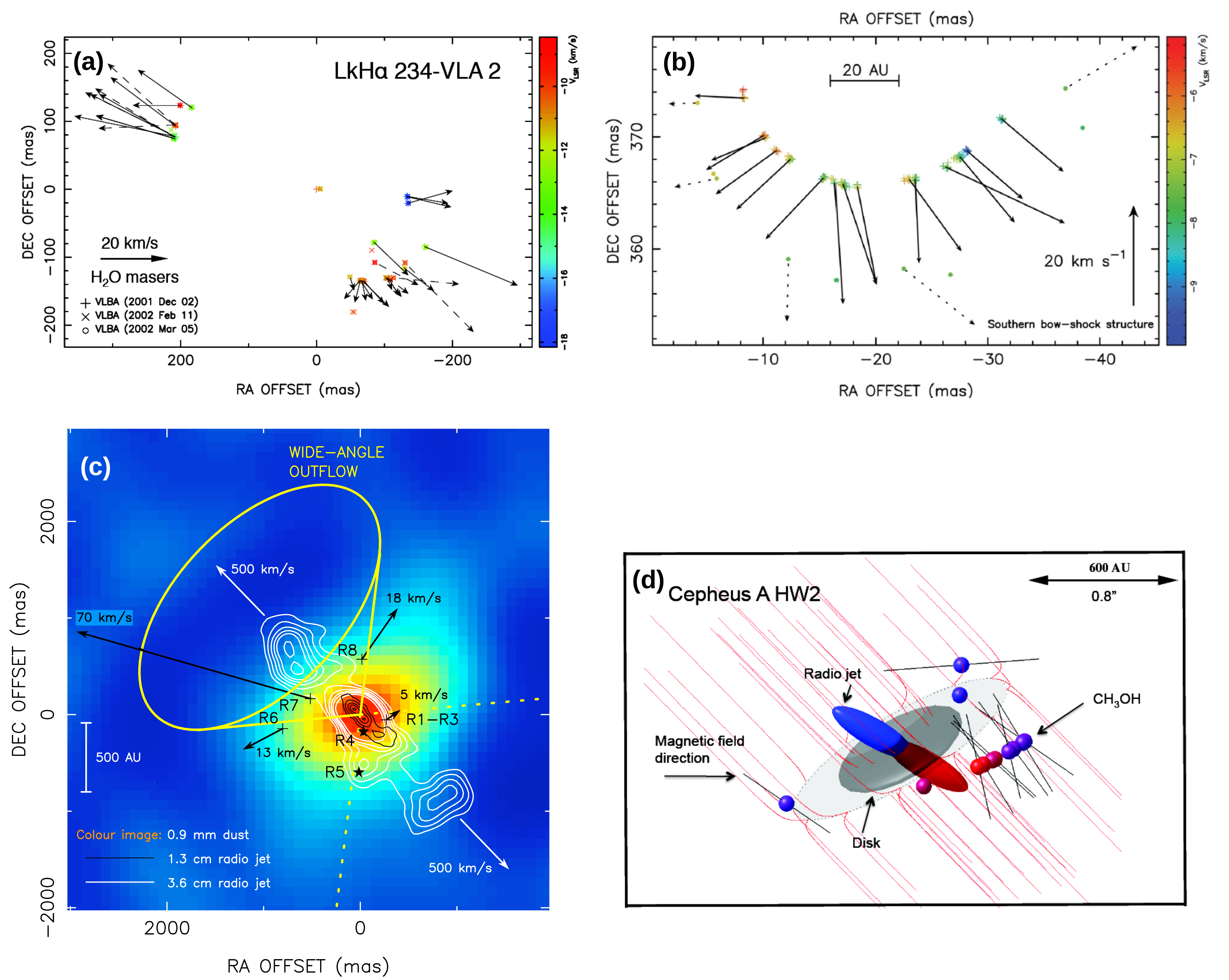}
 \caption{(a) Distribution and proper motions of H$_2$O masers showing a 
very compact ($\sim$180~au), short-lived ($\sim$40~yr), bipolar jet from 
a very embedded protostar of unknown nature in the LkH$\alpha$ 234 
star-forming region (Torrelles et al. 2014). (b) Micro-bow shock traced 
by water masers in AFGL2591 VLA 3-N. Arrows indicate the measured proper 
motions (Trinidad et al. 2013). (c) Maser microstructures tracing a 
low-collimation outflow around the radio jet associated with the massive 
protostar Cep A HW2 (Torrelles et al. 2011). (d) Magnetic field 
structure around the protostar-disk-jet system of Cep A HW2. Spheres 
indicate the CH$_3$OH masers and black vectors the magnetic field 
direction. Image adapted from Vlemmings et al.\ (2010).
 \label{fig:masers} 
}
\end{center}
\end{figure}

Molecular maser emission at cm wavelengths (e.g., H$_2$O, CH$_3$OH, OH) 
is often found in the early evolutionary stages of massive protostars. 
Such maser emission is usually very compact and strong, with brightness 
temperatures exceeding in some cases 10$^{10}$~K, allowing the 
observation of outflows at milliarcsecond (mas) scales (1 mas = 1 au at 
a distance of 1 kpc) using Very Long Baseline Interferometry (VLBI). 
Sensitive VLBI observations show, in some sources, thousands of maser 
spots forming microstructures that reveal the 3D kinematics of outflows 
and disks at small scale (e.g., Sanna et al. 2015). This kind of 
observations have given a number of interesting results: the discovery 
of short-lived, episodic non-collimated outflow events (e.g., Torrelles 
et al. 2001, 2003; Surcis et al. 2014); detection of infall motions in 
accretion disks around massive protostars (Sanna et al. 2017); the 
imaging of young ($< 100$ yr) small scale (a few 100 au) bipolar jets of 
masers (Sanna et al. 2012, Torrelles et al. 2014; 
Fig.~\ref{fig:masers}a), and even allowed to analyze the small-scale 
(1-20 au) structure of the micro bow-shocks (Uscanga et al. 2005; 
Trinidad et al. 2013; Fig.~\ref{fig:masers}b); the simultaneous presence 
of a wide-angle outflow and a collimated jet in a massive protostar 
(Torrelles et al. 2011; see Fig.~\ref{fig:masers}c); and polarization 
studies have determined the distribution and strength of the magnetic 
field very close to protostars allowing to better understand its role in 
the star formation processes (Surcis et al. 2009; Vlemmings et al. 2010; 
Sanna et al. 2015; Fig.~\ref{fig:masers}d).

In particular, the combination of continuum and maser studies in Cep A 
HW2 has been useful to provide one of the best examples of the two-wind 
model for outflows from massive protostars (Torrelles et al. 2011; 
Fig.~\ref{fig:masers}c). In this source the water masers trace the 
presence of a relatively slow ($\sim$10-70 km s$^{-1}$) wide-angle 
outflow (opening angle of $\sim102^\circ$), while the thermal jet traces 
a fast ($\sim$500 km s$^{-1}$) highly collimated radio jet (opening 
angle of $\sim18^\circ$). This two-wind phenomenon had been previously 
imaged only in low mass protostars such as L1551-IRS5 (Itoh et al. 2000, 
Pyo et al. 2005), HH 46/47 (Velusamy et al. 2007), HH 211 (Gueth \& 
Guilloteau 1999, Hirano et al. 2006) and IRAS 04166+2706 
(Santiago-Garc\'\i a et al. 2009). An extensive study of the connection 
between high velocity collimated jets and slow un-collimated winds in a 
large sample of low-mass class II objects has been carried out recently 
by Nisini et al. (2018); this work has been performed from the analysis 
of [OI]6300 \AA\ line profiles under the working hypothesis that the low 
velocity component traces a wide wind and the high velocity component a 
collimated jet.

\section{Nature of the Centimeter Continuum Emission in Thermal Radio 
Jets}

\subsection{Observational Properties: Radio to Bolometric Luminosity 
Correlation\label{sect:lbol}}

\begin{figure}[h]
\begin{center}
  \includegraphics[width=0.95\textwidth]{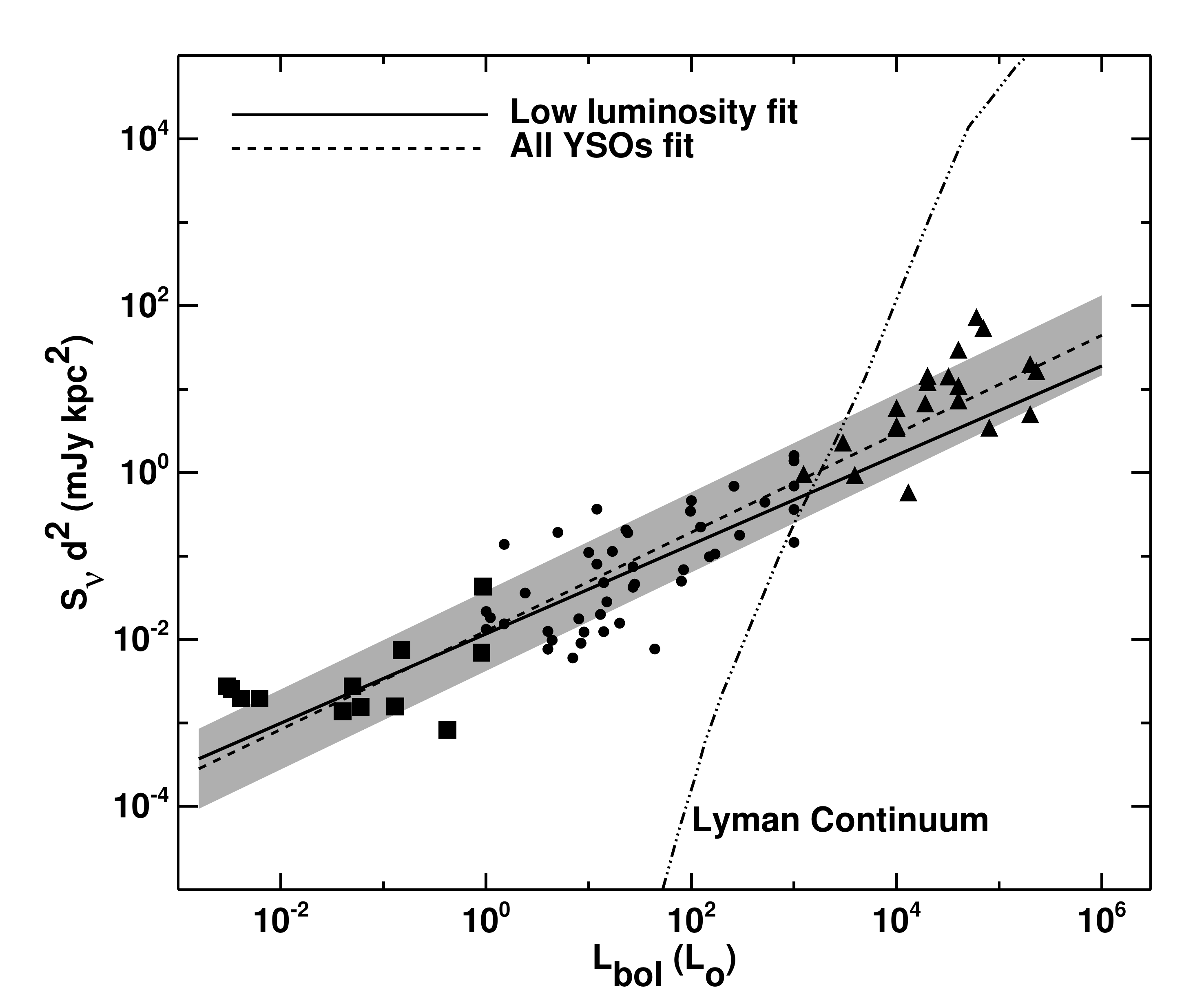}
 \caption{Empirical correlation between the bolometric luminosity and 
the radio continuum luminosity at cm wavelengths. Data are taken from 
Table \ref{tab:cm}. Triangles correspond to high luminosity objects 
($L_{\rm bol} > 1000~L_\odot$), dots correspond to low luminosity 
objects ($1 \le L_{\rm bol} \le 1000~L_\odot$), and squares to very low 
luminosity objects ($L_{\rm bol} < 1~L_\odot$). The dashed line is a 
least-squares fit to all the data points and the grey area indicates the 
residual standard deviation of the fit. The solid line is a fit to the 
low-luminosity objects alone. The dot-dashed line corresponds to the 
expected radio luminosity of an optically thin region photoionized by 
the Lyman continuum of the star.
 }
\label{fig:cor1}       
\end{center}
\end{figure}

Photoionization does not appear to be the ionizing mechanism of radio 
jets since, in the sources associated with low luminosity objects, the 
rate of ionizing UV photons ($\lambda<912$ \AA) from the star is clearly 
insufficient to produce the ionization required to account for the 
observed radio continuum emission (e.g., Rodr\'{\i}guez et al.\ 1989b; 
Anglada 1995). For low-luminosity objects ($1 \la L_{\rm bol} \la 
1000~L_\odot$), the observed flux densities at cm wavelengths are 
several orders of magnitude higher than those expected by 
photoionization (see Fig. \ref{fig:cor1} and left panel in Fig.~5 of 
Anglada 1995). Ionization by shocks in a strong stellar wind or in the 
jet itself has been proposed as the most likely possibility (Torrelles 
et al. 1985, Curiel et al. 1987, 1989, Anglada et al. 1992; see below). 
Detailed simulations of the two-shock internal working surfaces 
traveling down the jet flow, and the expected emission of the ionized 
material at shorter wavelengths have been performed (e.g., the H$\alpha$ 
and [OI]6300 \AA\ line emission of a radiative jet model with a variable 
ejection velocity by Raga et al. 2007).

\begin{table}[h!]
{\small 
\caption{Sample of YSOs Associated with Outflows and Radio Continuum Emission \label{tab:cm}}
\begin{tabular}{lccccl}
\hline\noalign{\smallskip}
  {Source}
& {$d$}
& {$L_{\rm bol}$}
& {$S_\nu$}
& {$\dot P$}
& {Refs.} \\
& {(kpc)}
& {($L_\odot$)}
& {(mJy)}
& {($M_\odot\,{\rm yr^{-1}\,km~s^{-1}}$)} \\
\noalign{\smallskip}\hline\noalign{\smallskip}
 J041757		& 0.140		& 0.003		& 0.14		& $\cdots$ 		& 79, 83, 84 \\	
 J041836		& 0.140		& 0.0033	& 0.13		& $\cdots$		& 84 	\\	
 J041847		& 0.140		& 0.0041	& 0.10		& $\cdots$		& 84	\\	
 J041938		& 0.140		& 0.0062 	& 0.10		& $\cdots$		& 84	\\	
 IRAM04191		& 0.140		& 0.05		& 0.14		& 1.5$\times10^{-5}$	& 70, 71, 72 \\	
 L1014IRS		& 0.260		& 0.15 		& 0.11 		& 1.3$\times10^{-7}$	& 70, 73, 74, 75 \\	
 L1148-IRS		& 0.140		& 0.13		& 0.080		& 1.0$\times10^{-7}$	& 70, 76, 77	\\	
 L1521F-IRS		& 0.140		& 0.04		& 0.07 		& 1.1$\times10^{-6}$	& 77, 78, 79	\\	
 IC348-SMM2E		& 0.240		& 0.06		& 0.027		& 1.1$\times10^{-7}$	& 79, 80, 81, 82 \\
 HH30			& 0.140		& 0.42 		& 0.042		& 7.5$\times10^{-7}$ 	& 85, 86, 87	\\
 L1262          	& 0.2 		& 1.0  		& 0.33		& 4.0$\times10^{-5}$ 	& 47, 50, 51, 142, 46 \\
 VLA1623        	& 0.16		& 1.5 		& 0.6 		& 4.5$\times10^{-4}$   	& 16, 35 	\\
 L723           	& 0.3 		& 2.4 		& 0.4 		& 3.2$\times10^{-4}$   	& 10, 11, 33, 28, 132, 155 \\
 L1489          	& 0.14 		& 4.4 		& 0.5 		& 2.0$\times10^{-7}$   	& 3, 4, 24, 137 \\
 B335           	& 0.25 		& 4 		& 0.2 		& 2.0$\times10^{-5}$   	& 10, 12, 24, 22, 58, 140, 142 \\
 NGC2264G       	& 0.8 		& 5 		& 0.3 		& 3.0$\times10^{-3}$   	& 17, 18, 36  	\\
 AS353A         	& 0.3 		& 8.4 		& 0.1 		& 4.5$\times10^{-4}$   	& 1, 23, 52, 12 \\
 L1448C         	& 0.35 		& 9 		& 0.1 		& 4.5$\times10^{-4}$   	& 14, 15, 34	\\
 L1448N(A)      	& 0.35 		& 10 		& 0.9 		& 1.2$\times10^{-4}$   	& 14, 15, 35, 129, 130, 140 \\
 RNO43          	& 0.4 		& 12 		& 0.5 		& 2.1$\times10^{-4}$  	& 12, 22, 39 	\\
 L483           	& 0.2 		& 14 		& 0.31 		& 1.5$\times10^{-5}$   	& 47, 45, 24	\\
 L1251B         	& 0.2 		& 14 		& 1.2 		& 4.9$\times10^{-5}$   	& 48, 142, 49	\\
 TTAU           	& 0.14 		& 17 		& 5.8 		& 4.1$\times10^{-5}$  	& 5, 6, 25, 26, 19, 1, 27 \\
 HH111          	& 0.46 		& 24 		& 0.9 		& 5.0$\times10^{-5}$  	& 20, 21, 23 	\\
 IRAS16293      	& 0.16 		& 27 		& 2.9 		& 1.6$\times10^{-3}$  	& 8, 9, 30, 31, 32	\\
 L1251A         	& 0.3 		& 27 		& 0.47 		& 8.6$\times10^{-5}$  	& 48, 45, 49	\\
 HARO4-255FIR   	& 0.48 		& 28 		& 0.2 		& 1.3$\times10^{-4}$  	& 12, 19, 37, 38 \\
 L1551-IRS5 		& 0.14 		& 20 		& 0.8 		& 1.0$\times10^{-4}$  	& 1, 6, 28, 101, 102, 103 \\
 HLTAU          	& 0.16 		& 44 		& 0.3 		& 1.4$\times10^{-6}$  	& 5, 19, 1, 29, 23, 7, 119 \\
 L1228          	& 0.20 		& 7 		& 0.15 		& 5.4$\times10^{-6}$  	& 53, 94, 198	\\
 PVCEP          	& 0.5 		& 80 		& 0.2 		& 1.0$\times10^{-4}$  	& 12, 13, 19, 27 \\
 L1641N         	& 0.42 		& 170 		& 0.6 		& 9.0$\times10^{-4}$  	& 43, 37, 44, 142 \\
 FIRSSE101      	& 0.45 		& 123 		& 1.1 		& 2.0$\times10^{-3}$  	& 37, 40, 41 	\\
 HH7-11~VLA3 		& 0.35 		& 150 		& 0.8 		& 1.1$\times10^{-3}$  	& 1, 2, 23, 69 	\\
 RE50           	& 0.46 		& 295 		& 0.84 		& 3.0$\times10^{-4}$  	& 42, 37, 40, 23, 41, 116 \\
 SERPENS        	& 0.415  	& 98  		& 2.0  		& 5.3$\times10^{-4}$    & 110, 111, 112, 156, 92, 93  	\\
 IRAS22198      	& 1.3 		& 1240		& 0.57 		& 2.3$\times10^{-4}$ 	& 88, 89 \\
 NGC2071-IRS3   	& 0.39 		& 520 		& 2.9 		& 1.5$\times10^{-2}$ 	& 54, 90, 91, 136   \\
 HH34			& 0.42 		& 15 		& 0.160 	& 3.0$\times10^{-6}$	& 59, 94, 95 	\\
 AF5142~CM1 		& 2.14 		& 10000 	& 1.3 		& 2.4$\times10^{-3}$ 	& 96, 108 	\\
 AF5142~CM2 		& 2.14 		& 1000 		& 0.35  	& 2.4$\times10^{-3}$ 	& 96, 108 	\\
 CepAHW2       		& 0.725 	& 10000 	& 6.9 		& 5.4$\times10^{-3}$ 	& 104, 105, 106, 107 	\\
 IRAS20126    		& 1.7 		& 13000 	& 0.2 		& 6.0$\times10^{-3}$ 	& 152, 153, 199 \\
 HH80-81      		& 1.7 		& 20000 	& 5.0 		& 1.0$\times10^{-3}$ 	& 98, 113, 114, 115 	\\
 V645Cyg     		& 3.5 		& 40000 	& 0.6 		& 7.0$\times10^{-4}$ 	& 122, 129, 137   \\
 IRAS16547    		& 2.9 		& 60000 	& 8.7 		& 4.0$\times10^{-1}$ 	& 149, 150, 151 \\
 IRAS18566~B 		& 6.7 		& 80000 	& 0.077  	& 7.2$\times10^{-3}$	& 99, 100, 167   \\
 IRAS04579 		& 2.5 		& 3910 		& 0.15 		& 9.0$\times10^{-4}$ 	& 88, 148, 194	\\
 GGD14-VLA7     	& 0.9  		& 1000 		& 0.18 		& 1.6$\times10^{-3}$   	& 55, 65, 66, 67, 68, 97 \\
\noalign{\smallskip}\hline
\end{tabular}
}
\end{table}

\setcounter{table}{1}
\begin{table}[h!]
{\small 
\caption{(Continued) \label{tab:cmnew2}}
\begin{tabular}{lccccl}
\hline\noalign{\smallskip}
  {Source}
& {$d$}
&  {$L_{\rm bol}$}
&  {$S_\nu$}
&  {$\dot P$}
& {Refs.} \\
& {(kpc)}
& {($L_\odot$)}
& {(mJy)}
& {($M_\odot\,{\rm yr^{-1}\,km~s^{-1}}$)} \\
\noalign{\smallskip}\hline\noalign{\smallskip}
 IRAS23139     		& 4.8 		& 20000 	& 0.53 		& 8.0$\times10^{-4}$ 	& 123, 124, 125 \\
 N7538-IRS9 		& 2.8 		& 40000 	& 3.8 		& 2.1$\times10^{-2}$ 	& 88, 126, 127 	\\
 N7538-IRS9~A1  	& 2.8 		& 40000 	& 1.4 		& 3.0$\times10^{-2}$ 	& 88, 126, 127 	\\
 IRAS16562      	& 1.6 		& 70000 	& 21.1 		& 3.0$\times10^{-2}$ 	& 131, 133 	\\
 I18264-1152~F  	& 3.5 		& 10000 	& 0.28 		& 1.4$\times10^{-2}$ 	& 99, 100, 123  \\
 G31.41         	& 7.9 		& 200000 	& 0.32 		& 6.0$\times10^{-2}$ 	& 134, 135, 143, 147 	\\
 AF2591-VLA3    	& 3.3 		& 230000 	& 1.52  	& 7.7$\times10^{-3}$ 	& 121, 172, 195  	\\
 I18182-1433b   	& 4.5 		& 20000 	& 0.6 		& 2.8$\times10^{-3}$ 	& 99, 196  	\\
 I18089-1732(1)a 	& 3.6 		& 32000 	& 1.1 		& $\cdots$    		& 196 	 \\
 G28S-JVLA1     	& 4.8 		& 100 		& 0.02 		& $\cdots$    		& 197 	\\
 G28N-JVLA2N    	& 4.8 		& 1000 		& 0.06 		& $\cdots$    		& 197 	\\
 G28N-JVLA2S    	& 4.8 		& 1000 		& 0.03 		& $\cdots$    		& 197 	\\
 L1287 			& 0.85		& 1000 		& 0.5 		& 3.3$\times10^{-4}$	& 192, 193, 116, 118 	\\
 DG~Tau~B 		& 0.15		& 0.9 		& 0.31		& 6.0$\times10^{-6}$	& 60, 61, 157, 158 	\\
 Z~CMa 			& 1.15		& 3000 		& 1.74		& 1.0$\times10^{-4}$ 	& 122, 159, 160	\\
 W3IRS5(d) 		& 1.83 		& 200000 	& 1.5 		& 3.0$\times10^{-2}$ 	& 128, 161, 162 	\\
 YLW~16A 		& 0.16 		& 13 		& 0.78 		& 3.4$\times10^{-6}$ 	& 137, 163  	\\
 YLW~15~VLA1 		& 0.12 		& 1		& 1.5 		& 1.1$\times10^{-5}$ 	& 138, 164, 165, 166	\\
 L778 			& 0.25 		& 0.93 		& 0.69 		& 1.9$\times10^{-6}$ 	& 4, 137, 168 	 \\
 W75N(B)VLA1 		& 1.3  		& 19000 	& 4.0 		& 1.2$\times10^{-2}$ 	& 139, 169, 170, 171, 172, 154  \\
 NGC1333~VLA2 		& 0.235		& 1.5 		& 2.5  		& 4.9$\times10^{-4}$ 	& 140, 141, 173, 174, 175 	\\
 NGC1333~IRAS4A1	& 0.235  	& 8 		& 0.32 		& 1.5$\times10^{-4}$ 	& 140, 173, 176, 177, 178 	\\
 NGC1333~IRAS4B 	& 0.235 	& 1.1 		& 0.33 		& 4.1$\times10^{-5}$ 	& 140, 173, 177, 178 	\\
 L1551~NE-A 		& 0.14 		& 4.0 		& 0.39 		& 1.5$\times10^{-5}$ 	& 140, 179, 180, 181, 61, 62 	\\
 Haro 6-10 VLA1 	& 0.14 		& 0.5  		& 1.1 		& 9.4$\times10^{-6}$  	& 146, 12, 182, 183, 184  \\
 L1527 VLA1 		& 0.14 		& 1.9 		& 1.1  		& 1.6$\times10^{-4}$ 	& 144, 145, 146, 179, 184 	\\
 OMC 2/3 VLA4 		& 0.414		& 40 		& 0.83  	& 5.2$\times10^{-4}$ 	& 146, 185 	\\
 HH1-2~VLA1 		& 0.414 	& 23 		& 1.2 		& 4.6$\times10^{-5}$ 	& 186, 187, 188, 109 	\\
 HH1-2~VLA3 		& 0.414 	& 84 		& 0.40 		& 5.7$\times10^{-4}$ 	& 186, 187, 188	\\
 OMC2 VLA11 		& 0.414 	& 360 		& 2.16 		& 3.2$\times10^{-3}$ 	& 189, 190, 191 \\
 HH46/47        	& 0.45 		& 12 		& 1.8  		& 2.0$\times10^{-4}$ 	& 23, 56, 57, 117, 120 	\\
 IRAS20050 		& 0.7 		& 260 		& 1.4 		& 5.0$\times10^{-3}$ 	& 63, 64 \\
\noalign{\smallskip}\hline
\end{tabular} \\
\vskip0.2cm
 References: {
(1) Edwards \& Snell 1984;
(2) Bachiller \& Cernicharo 1990;
(3) Rodr\'{\i}guez et al.\ 1989b;
(4) Myers et al.\ 1988;
(5) Calvet et al.\ 1983;
(6) Cohen et al.\ 1982;
(7) Brown et al.\ 1985;
(8) Wootten \& Loren 1987;
(9) Estalella et al.\ 1991;
(10) Goldsmith et al.\ 1984;
(11) Anglada et al.\ 1991;
(12) Anglada et al.\ 1992
(13) Levreault 1984;
(14) Bachiller et al.\ 1990;
(15) Curiel et al.\ 1990;
(16) Andr\'e et al.\ 1990;
(17) Margulis et al.\ 1988;
(18) G\'omez et al.\ 1994;
(19) Levreault 1988;
(20) Reipurth \& Olberg 1991;
(21) Rodr\'{\i}guez \& Reipurth 1994;
(22) Cabrit, Goldsmith, \& Snell 1988;
(23) Reipurth et al.\ 1993;
(24) Ladd et al.\ 1991;
(25) Schwartz et al.\ 1986;
(26) Edwards \& Snell 1982;
(27) Evans et al.\ 1986;
(28) Mozurkewich et al.\ 1986;
(29) Torrelles et al.\ 1987;
(30) Mundy et al.\ 1986;
(31) Mizuno et al.\ 1990;
(32) Walker et al.\ 1988;
(33) Avery et al.\ 1990;
(34) Bachiller et al.\ 1991;
(35) Andr\'e et al.\ 1993;
(36) Ward-Thompson et al.\ 1995;
(37) Morgan et al.\ 1991;
(38) Morgan \& Bally 1991;
(39) Cohen \& Schwartz 1987;
(40) Fukui 1989;
(41) Morgan et al.\ 1990;
(42) Reipurth \& Bally 1986;
(43) Fukui et al.\ 1988;
(44) Chen, H. et al.\ 1995;
(45) Beltr\'an et al.\ 2001;
(46) Terebey et al.\ 1989;
(47) Parker et al.\ 1988;
(48) Sato et al.\ 1994;
(49) Kun \& Prusti 1993;
(50) Yun \& Clemens 1994;
(51) Parker 1991;
(52) Cohen \& Bieging 1986;
(53) Haikala \& Laurenjis 1989;
(54) Snell et al.\ 1984;
(55) Rodr\'{\i}guez et al.\ 1982; 
(56) Chernin \& Masson 1991;
(57) S. Curiel, private com.;
(58) Moriarty-Schieven \& Snell 1989;
(59) Antoniucci et al.\ 2008;
(60) Mitchell et al.\ 1994;
(61) Rodr\'{\i}guez et al.\ 1995;
(62) Moriarty-Schieven et al.\ 1995;
(63) Bachiller et al.\ 1995;
(64) Wilking et al.\ 1989;
(65) Little et al.\ 1990; 
(66) G\'omez et al.\ 2000; 
(67) G\'omez et al.\ 2002; 
(68) Harvey et al.\ 1985; 
(69) Rodr\'{\i}guez et al.\ 1997; 
(70) Dunham et al.\ 2008; 
(71) Choi et al.\ 2014; 
(72) Andr\'e et al.\ 1999; 
}
}
\end{table}

\setcounter{table}{1}
\begin{table}[h!]
{\small 
\caption{(Continued) \label{tab:cmnew3}}
\begin{tabular}{lccccl}
\\
\end{tabular}
{
\noindent

(73) Bourke et al.\ 2005; 
(74) Shirley et al.\ 2007; 
(75) Maheswar et al.\ 2011; 
(76) Kauffmann et al.\ 2011; 
(77) AMI Consortium et al.\ 2011a; 
(78) Takahashi et al.\ 2013; 
(79) Palau et al.\ 2014;
(80) Rodr\'{\i}guez et al.\ 2014b; 
(81) Hirota et al.\ 2008; 
(82) Hirota et al.\ 2011; 
(83) Palau et al.\ 2012; 
(84) Morata et al.\ 2015;  
(85) Cotera et al.\ 2001; 
(86) Pety et al.\ 2006; 
(87) G. Anglada et al., in prep.; 
(88) S\'anchez-Monge et al.\ 2008; 
(89) Zhang et al.\ 2005; 
(90) Carrasco-Gonz\'alez et al.\ 2012a; 
(91) Stojimirovi\'c et al.\ 2008; 
(92) Kristensen et al.\ 2012; 
(93) van Kempen et al.\ 2016; 
(94) Rodr\'{\i}guez \& Reipurth 1996; 
(95) Chernin \& Masson 1995,; 
(96) Zhang et al.\ 2007; 
(97) Dzib et al.\ 2016; 
(98) Benedettini et al.\ 2004; 
(99) Beuther et al.\ 2002; 
(100) Rosero et al.\ 2016;
(101) Rodr\'{\i}guez et al.\ 2003b; 
(102) Bieging \& Cohen 1985; 
(103) Rodr\'{\i}guez et al.\ 1986; 
(104) Hughes 1988;  
(105) Rodr\'{\i}guez et al.\ 1994b; 
(106) G\'omez et al.\ 1999;  
(107) Curiel et al.\ 2006; 
(108) Hunter et al.\ 1995;  
(109) Rodr\'{\i}guez et al.\ 1990; 
(110) Rodr\'{\i}guez et al.\ 1989a; 
(111) Curiel et al.\ 1993; 
(112) Curiel 1995; 
(113) Rodr\'{\i}guez \& Reipurth 1989; 
(114) Mart\'{\i} et al.\ 1993; 
(115) Mart\'{\i} et al.\ 1995; 
(116) Anglada 1995; 
(117) Arce et al.\ 2013; 
(118) Anglada et al.\ 1994;  
(119) Rodr\'{\i}guez et al.\ 1994a; 
(120) Zhang et al.\ 2016; 
(121) Hasegawa \& Mitchell 1995; 
(122) Skinner et al.\ 1993; 
(123) Sridharan et al.\ 2002; 
(124) Wouterloot et al.\ 1989; 
(125) Trinidad et al.\ 2006; 
(126) Tamura et al.\ 1991; 
(127) Sandell et al.\ 2005; 
(128) Claussen et al.\ 1994; 
(129) Girart et al.\ 1996a;  
(130) Girart et al.\ 1996b; 
(131) Guzm\'an et al.\ 2010; 
(132) Anglada et al.\ 1996; 
(133) Guzm\'an et al.\ 2011; 
(134) Osorio et al.\ 2009; 
(135) Cesaroni et al.\ 2010; 
(136) Torrelles et al.\ 1998; 
(137) Girart et al.\ 2002;  
(138) Girart et al.\ 2000; 
(139) Torrelles et al.\ 1997; 
(140) Reipurth et al.\ 2002; 
(141) Rodr\'{\i}guez et al.\ 1999;  
(142) Anglada et al.\ 1998;
(143) Cesaroni et al.\ 2011; 
(144) Rodr\'{\i}guez \& Reipurth 1998; 
(145) Loinard et al.\ 2002; 
(146) Reipurth et al.\ 2004;  
(147) Mayen-Gijon 2015; 
(148) Molinari et al.\ 1996;
(149) Garay et al.\  2003; 
(150) Brooks et al.\ 2003;  
(151) Rodr\'{\i}guez et al.\ 2005; 
(152) Hofner et al.\ 1999; 
(153) Trinidad et al.\ 2005; 
(154) Torrelles et al.\ 2003; 
(155) Carrasco-Gonz\'alez et al.\ 2008b; 
(156) Rodr\'{\i}guez-Kamenetzky et al.\ 2016; 
(157) Rodr{\'{\i}}guez et al.\ 2012b; 
(158) Zapata et al.\ 2015; 
(159) Evans et al.\ 1994; 
(160) Vel{\'a}zquez \& Rodr{\'{\i}}guez 2001; 
(161) Imai et al.\ 2000; 
(162) Hasegawa et al.\ 1994; 
(163) Sekimoto et al.\ 1997; 
(164) Girart et al.\ 2004; 
(165) van Kempen et al.\ 2009; 
(166) Bontemps et al.\ 1996; 
(167) Hofner et al.\ 2017; 
(168) Rodr{\'{\i}}guez et al.\ 1989b;
(169) Hunter et al.\ 1994; 
(170) Shepherd et al.\ 2003; 
(171) Carrasco-Gonz{\'a}lez et al.\ 2010a; 
(172) Rygl et al.\ 2012; 
(173) Plunkett et al.\ 2013; 
(174) Sadavoy et al.\ 2014; 
(175) Downes \& Cabrit 2007; 
(176) Dunham et al.\ 2014; 
(177) Ching et al.\ 2016; 
(178) Y{\i}ld{\i}z et al.\ 2012; 
(179) Froebrich 2005; 
(180) Lim et al.\ 2016; 
(181) Takakuwa et al.\ 2017; 
(182) Doppmann et al.\ 2008; 
(183) Roccatagliata et al.\ 2011; 
(184) Hogerheijde et al.\ 1998; 
(185) Yu et al.\ 2000; 
(186) Fischer et al.\ 2010; 
(187) Rodr{\'{\i}}guez et al.\ 2000; 
(188) Correia et al.\ 1997; 
(189) Furlan et al.\ 2016; 
(190) Osorio et al.\ 2017; 
(191) Takahashi et al.\ 2008; 
(192) Yang et al.\ 1991; 
(193) Wu et al.\ 2010; 
(194) Xu et al.\ 2012; 
(195) Johnston et al.\ 2013; 
(196) Zapata et al.\ 2006; 
(197) C. Carrasco-Gonz\'alez et al., in prep;
(198) Skinner et al.\ 2014;
(199) Shepherd et al.\ 2000.
}
}
\end{table}

In Figure \ref{fig:cor1} we plot the observed cm luminosity ($S_\nu 
d^2$) as a function of the bolometric luminosity for the sources listed 
in Table \ref{tab:cm} (squares, dots, and triangles correspond 
respectively to very low, low, and high luminosity objects). For most of 
the sources in this plot, $S_\nu$ is the flux density at 3.6 cm, but 
some data at 6, 2, and 1.3 cm are included to construct a larger sample; 
these data points are used to estimate the 3.6 cm luminosity using the 
spectral index, when known, or assuming a value of $\sim$0.5. As can be 
seen in Figure \ref{fig:cor1}, the observed cm luminosity (data points) 
is uncorrelated with the cm luminosity expected from photoionization 
(dot-dashed line), further indicating that this is not the ionizing 
mechanism. However, as the figure shows, the observed cm luminosity is 
indeed correlated with the bolometric luminosity ($L_{\rm bol}$). A fit 
to the 48 data points with $1~L_\odot \la L_{\rm bol} \la 1000~L_\odot$ 
(dots) gives:
 \begin{equation} 
 {\biggl(\frac{S_\nu d^2}{\rm mJy~kpc^2}\biggr)} = 10^{-1.93\pm0.14}
\biggl(\frac{L_{\rm bol}}{L_\odot}\biggr)^{0.54\pm0.08}.
 \end{equation}

As can be seen in the figure, the correlation also holds for both the 
most luminous (triangles) and the very low luminosity (squares) objects, 
suggesting that the mechanism that relates the bolometric and cm 
luminosities is shared by all the YSOs.
 A fit to all the 81 data points ($10^{-2}~L_\odot \la L_{\rm bol} \la 
10^6~L_\odot$) gives a better fit, with a similar result,
 \begin{equation} \label{eq:lbol}
 {\biggl(\frac{S_\nu d^2}{\rm mJy~kpc^2}\biggr)} = 10^{-1.90\pm0.07} 
\biggl(\frac{L_{\rm bol}}{L_\odot}\biggr)^{0.59\pm0.03}.
 \end{equation}

A correlation between bolometric and cm luminosities was noted by Cabrit 
\& Bertout (1992) from a set of $\sim$25 outflow sources, quoting a 
slope of $\sim$0.8 in a log-log plot. Skinner et al. (1993) suggested a 
correlation between the 3.6 cm luminosity and the bolometric luminosity 
with a slope of 0.9 by fitting a sample of Herbig Ae/Be stars and 
candidates with 11 detections and 7 upper limits. The fit to the 29 
outflow sources presented in Anglada (1995) gives a slope of 0.6 (after 
updating some distances and flux densities; a slope of 0.7 was obtained 
with the values originally listed in Table 1 of that paper), similar to 
the value given in equation~(\ref{eq:lbol}). Shirley et al. (2007) 
obtained separate fits for the 3.6 cm and for the 6 cm data, obtaining 
slopes of 0.7 and 0.9, respectively (although these fits are probably 
affected by a few outliers with anomalously high values of the flux 
density, taken from the compilation of Furuya et al. 2003). Because 
radio jets have positive spectral indices with typical values around 
$\sim0.4$, it is expected that the 3.6 cm flux densities are $\sim$20\% 
higher than the 6 cm flux densities, but the slopes of the luminosity 
correlations are expected to be similar, provided the scattering in the 
values of the spectral index is small. L. Tychoniec et al. (in prep.) 
analyze a large homogeneous sample of low-mass protostars in Perseus, 
obtaining similar slopes of $\sim$0.7 for data at 4.1 and 6.4 cm. These 
authors note, however, that the correlations are weak for these sources 
that cover a relatively small range of luminosities. These results 
suggest that there is a general trend over a wide range of luminosities, 
but with an intrinsic dispersion.

Recently, Moscadelli et al. (2016) derived a slope of 0.5 from a small 
sample (8 sources) of high luminosity objects with outflows traced by 
masers. Also, recent surveys targeted towards high-mass protostellar 
candidates (but without a confirmed association with an outflow) also 
show radio continuum to bolometric luminosity correlations with similar 
slopes of $\sim$0.7 (Purser et al. 2016). Recently, Tanaka et al. (2016) 
modeled the evolution of a massive protostar and its associated jet as 
it is being photoionized by the protostar, making predictions for the 
free-free continuum and RRL emissions. These authors find global 
properties of the continuum emission similar to those of radio jets. The 
radio continuum luminosities of the photoionized outflows predicted by 
the models are somewhat higher than those obtained from the empirical 
correlations for jets but much lower than those expected for optically 
thin HII regions. When including an estimate of the ionization of the 
ambient clump by photons that escape along the outflow cavity the 
predicted properties get closer to those of the observed UC/HC HII 
regions. Photoevaporation has not been included. This kind of models, 
including additional effects such as the photoevaporation, are a 
promising tool to investigate the transition from jets to HII regions in 
massive protostars. Purser et al. (2016) found a number of objects with 
radio luminosities intermediate between optically thin HII regions and 
radio jets that these authors interpret as optically thick HII regions. 
We note that this kind of objects could be in an evolutionary stage 
intermediate between the jet and the HII region phases, and their 
properties could be predicted by models similar to those of Tanaka et 
al. (2016).

From a sample of outflow sources of low and very low bolometric 
luminosity, but using low angular resolution data ($\sim30''$) at 1.8 
cm, a correlation between the cm luminosity and the internal 
luminosity\footnote{The internal luminosity is the luminosity in excess 
of that supplied by the interstellar radiation field.} with a slope of 
either 0.5 or 0.6 was found (AMI Consortium et al. 2011a, b), depending, 
respectively, on the use of either the bolometric luminosity or the IR 
luminosity as an estimate of the internal luminosity.  Morata et al. 
(2015) report cm emission from four proposed proto-brown dwarf 
(proto-BD) candidates, that appear to follow the general trend of the 
luminosity correlation but showing some excess of radio emission. 
Further observations are required to confirm the nature of these objects 
as proto-BDs, to better determine their properties such as distance and 
intrinsic luminosity, as well as their radio jet morphology. If 
confirmed as bona fide proto-BDs that follow the correlation, this would 
suggest that the same mechanisms are at work for YSOs and proto-BDs, 
supporting the idea that the intrinsic properties of proto-BDs are a 
continuation to smaller masses of the properties of low-mass YSOs. It is 
interesting that the only two \sl bona fide\rm~ young brown dwarfs 
detected in the radio continuum fall well in this correlation (Rodr\'\i 
guez et al. 2017).

Recently, it has been realized that young stars in more advanced stages, 
such as those surrounded by transitional disks\footnote{Transitional 
disks are accretion disks with central cavities or gaps in the dust 
distribution that are attributed to disk clearing by still forming 
planets.} are also associated with radio jets that have become 
detectable with the improved sensitivity of the JVLA (Rodr\'\i guez et 
al. 2014a, Mac\'\i as et al. 2016). In these objects, accretion is very 
low but high enough to produce outflow activity detectable through the 
associated radio emission at a level of $\lesssim$0.1 mJy. These radio 
jets follow the general trend of the luminosity correlation but appear 
to be radio underluminous with respect to the correlation that was 
established from data corresponding to younger objects. This fact has 
been interpreted as indicating that it is the accretion component of the 
luminosity that is correlated with the outflow (and, thus, with the 
radio flux of the jet). Accretion luminosity is dominant in the youngest 
objects, from which the correlation was derived, while in more evolved 
objects the stellar contribution (which is not expected to be correlated 
with the radio emission) to the total luminosity becomes more important. 
Thus, it is expected that in more evolved objects the observed radio 
luminosity is correlated with only a fraction of the bolometric 
luminosity.

In summary, the radio luminosity ($S_\nu d^2$) of thermal jets 
associated with very young stellar objects is correlated with their 
bolometric luminosity ($L_{\rm bol}$). This correlation is valid for 
objects of a wide range of luminosities, from high to very low 
luminosity objects, and likely even for proto-BDs. This suggests that 
accretion and outflow processes work in a similar way for objects of a 
wide range of masses and luminosities. Accretion appears to be 
correlated with outflow (which is traced by the radio luminosity). As 
the young star evolves, accretion decreases and so does the radio 
luminosity. However, for these more evolved objects the accretion 
luminosity represents a smaller fraction of the bolometric luminosity 
(the luminosity of the star becomes more important) and they become 
radio underluminous with respect to the empirical correlation, which was 
derived for younger objects.

\subsection{Observational Properties: Radio Luminosity to Outflow 
Momentum Rate Correlation \label{sect:pp}}

\begin{figure}[h]
\begin{center}
  \includegraphics[width=0.95\textwidth]{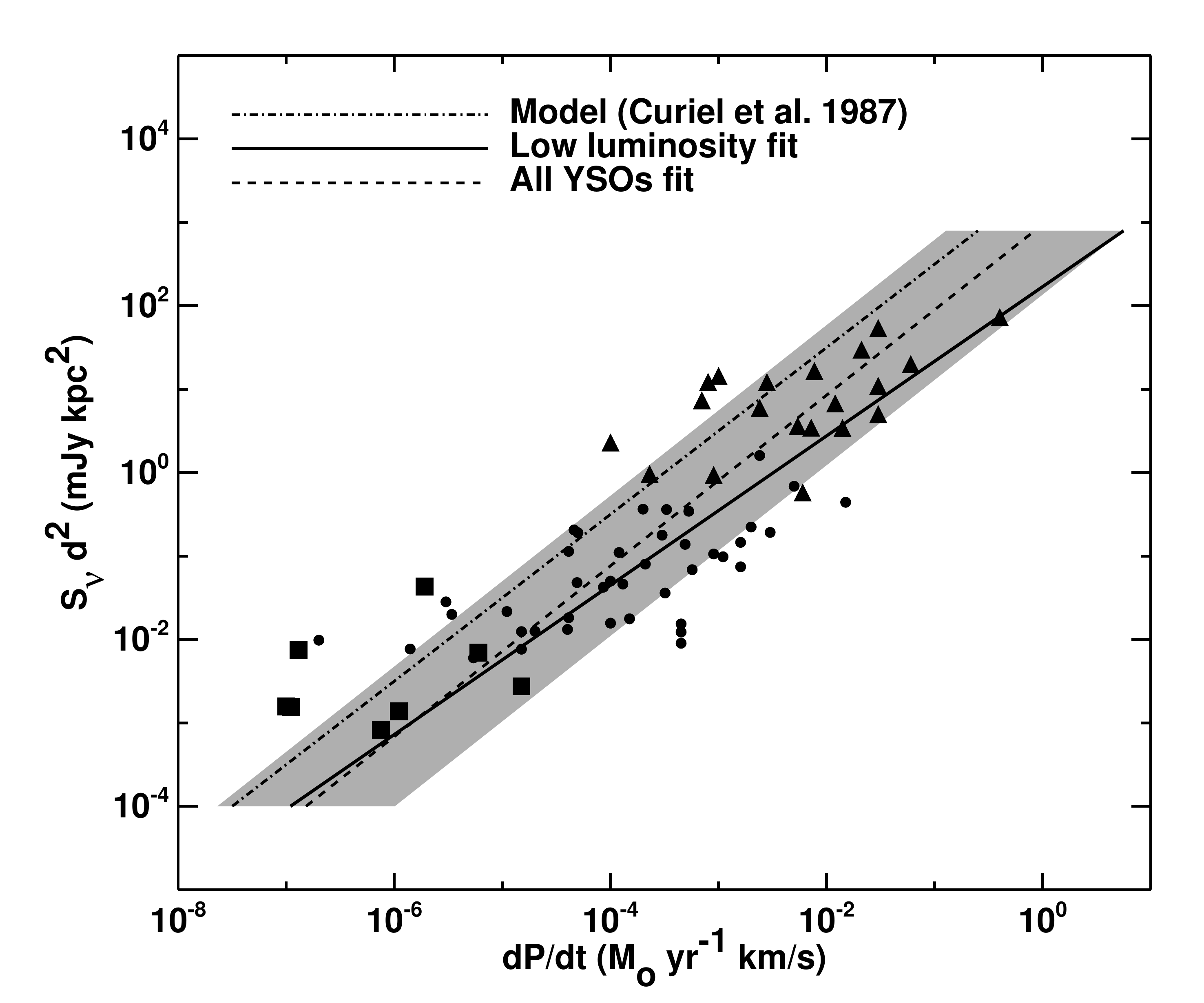}
 \caption{Empirical correlation between the outflow momentum rate and 
the radio continuum luminosity at cm wavelengths. Data are taken from 
Table \ref{tab:cm}. Triangles correspond to high luminosity objects 
($L_{\rm bol} > 1000~L_\odot$), dots correspond to low luminosity 
objects ($1 \le L_{\rm bol} \le 1000~L_\odot$), and squares to very low 
luminosity objects ($L_{\rm bol} < 1~L_\odot$). The dashed line is a 
least-squares fit to all the data points and the grey area indicates the 
residual standard deviation of the fit. The solid line is a fit to the 
low-luminosity objects alone. The dot-dashed line corresponds to the 
radio luminosity predicted by the models of Curiel et al. (1987, 1989).
}
 \label{fig:cor2}       
\end{center}
\end{figure}

The radio luminosity of thermal radio jets is also correlated with the 
properties of the associated outflows. A correlation between the 
momentum rate (force) in the outflow, $\dot P$, derived from 
observations, and the observed radio continuum luminosity at centimeter 
wavelengths, $S_{\nu}d^2$, was first noted by Anglada et al. (1992) and 
by Cabrit \& Bertout (1992). Anglada et al. (1992) considered a sample 
of 16 sources of low bolometric luminosity (to avoid a contribution from 
photoionization) and found a correlation $(S_\nu d^2/{\rm mJy~kpc^2}) = 
10^{2.4\pm1.0}\,(\dot P/M_{\sun}~{\rm yr^{-1}})^{0.9\pm0.3}$. Since the 
outflow momentum rate estimates have considerable uncertainties (more 
than one order of magnitude, typically), $\dot P$ was fitted (in the 
log-log space) taking $S_{\nu}d^2$ as the independent variable, as it 
was considered to be less affected by observational uncertainties. The 
correlation was confirmed with fits to larger samples (Anglada 1995, 
1996; Anglada et al. 1998; Shirley et al. 2007; AMI Consortium et al.\ 
2011a, b, 2012). The best fit to the low luminosity sources ($1 \la 
L_{\rm bol} \la 1000~L_\odot$) presented in Table \ref{tab:cm} gives 
(see also Fig. \ref{fig:cor2}):
 \begin{equation} 
{\biggl(\frac{S_\nu d^2}{\rm mJy~kpc^2}\biggr)} = 
10^{2.22\pm0.46} \biggl(\frac{\dot P}{M_\odot~{\rm
yr^{-1}~km~s^{-1}}}\biggr)^{0.89\pm0.16}.
 \label{eq:dotp1}
 \end{equation}

This correlation between the outflow momentum rate and the radio 
luminosity has been interpreted as evidence that shocks are the ionizing 
mechanism of jets. Curiel et al.\ (1987, 1989) modeled the scenario in 
which a neutral stellar wind is ionized as a result of a shock against 
the surrounding high-density material, assuming a plane-parallel shock.  
From the results obtained in this model, ignoring further details about 
the radiative transfer and geometry of the emitting region, and assuming 
that the free-free emission is optically thin, a relationship between 
the momentum rate in the outflow and the centimeter luminosity can be 
obtained (see Anglada 1996, Anglada et al. 1998):
 \begin{equation}
 \left(\frac{S_{\nu}d^2}{{\rm mJy~kpc^2}}\right) = 
 10^{3.5}\,\eta 
 \left(\frac{\dot{P}}{M_{\sun}~{\rm yr^{-1}}~\mbox{\kms}} \right),  
 \label{eq:curiel}
 \end{equation}
 where $\eta = \Omega/{4\pi}$ is an efficiency factor that can be taken 
to equal the fraction of the stellar wind that is shocked and produces 
the observed radio continuum emission. Despite the limitations of the 
model, and the simplicity of the assumptions used to derive 
equation~(\ref{eq:curiel}), its predictions agree quite well with the 
results obtained from a large number of observations (equation 
\ref{eq:dotp1}), for an efficiency $\eta \simeq 0.1$. Gonz\'alez \& 
Cant\'o (2002) present a model in which the ionization is produced by 
internal shocks in a wind, resulting of periodic variations of the 
velocity of the wind at injection.

Cabrit \& Bertout (1992) and Rodr\'\i guez et al. (2008) noted that 
high-mass protostars driving molecular outflows appear to follow the 
same radio luminosity to outflow momentum rate correlation as the 
sources of low luminosity. As can be seen in Figure \ref{fig:cor2}, high 
luminosity objects fall close to the fit determined by the low 
luminosity objects. Actually, a fit including all the sources in the 
sample of Table \ref{tab:cm}, including both low and high luminosity 
objects, gives a quite similar result,
 \begin{equation} 
{\biggl(\frac{S_\nu d^2}{\rm mJy~kpc^2}\biggr)} = 
10^{2.97\pm0.27} \biggl(\frac{\dot P}{M_\odot~{\rm
yr^{-1}~km~s^{-1}}}\biggr)^{1.02\pm0.08},
 \label{eq:dotp}
 \end{equation} as it was also the case for the bolometric luminosity 
correlation (Fig \ref{fig:cor1}). Thus, radio jets in massive protostars 
appear to be ionized by a mechanism similar to that acting in low 
luminosity objects. Radio jets would represent a stage in massive star 
formation previous to the onset of photoionization and the development 
of an HII region. Actually, the correlations can be used as a diagnostic 
tool to discriminate between photoionized (HII regions) versus 
shock-ionized (jets) sources (see Tanaka et al. 2016 and the previous 
discussion in Sect.~\ref{sect:lbol}).

As in the radio to bolometric luminosity correlation, the very low 
luminosity objects also fall in the outflow momentum rate correlation. 
These results are interpreted as indicating that the mechanisms for 
accretion, ejection and ionization of outflows are very similar for all 
kind of YSOs, from very low to high luminosity protostars.

A direct measure of the outflow momentum rate is difficult for objects 
in the last stages of the star formation process, when accretion has 
decreased to very small values and outflows are hard to detect. For 
these objects, the weak radio continuum emission can be used as a tracer 
of the outflow. Recent results obtained for sources associated with 
transitional disks (Rodr\'\i guez et al. 2014a, Mac\'\i as et al. 2016) 
indicate that the observed radio luminosities are consistent with the 
outflow momentum rate to radio luminosity correlation being valid and 
the ratio between accretion, and outflow being similar in these low 
accretion objects than in younger protostars.

\subsection{On the Origin of the Correlations \label{sect:origin}}

As has been shown above, the observable properties of the cm continuum 
outflow sources indicate that these sources trace thermal free-free 
emission from ionized collimated outflows (jets).  Both theoretical and 
observational results suggest that the ionization in thermal jets is 
only partial ($\sim$1-10\%; Rodr\'\i guez et al.\ 1990, Hartigan et al.\ 
1994, Bacciotti et al.\ 1995). The mechanism that is able to produce the 
required ionization, even at these relatively low levels, is still not 
fully understood. As photoionization cannot account for the observed 
radio continuum emission of low luminosity objects (see above), shock 
ionization has been proposed as a viable alternative mechanism (Curiel 
et al. 1987; Gonz\'alez \& Cant\'o 2002).

The correlations described in Sections \ref{sect:lbol} and \ref{sect:pp} 
are related to the well-known correlation between the momentum rate 
observed in molecular outflows and the bolometric luminosity of the 
driving sources, first noted by Rodr\'\i guez et al. (1982). More recent 
determinations of this correlation give $(\dot P/M_{\sun}~{\rm 
yr^{-1}~km~s^{-1}})$ = $10^{-4.36\pm0.12}$ $(L_{\rm 
bol}/L_\odot)^{0.69\pm0.05}$ (Cabrit \& Bertout 1992) or $(\dot 
P/M_{\sun}~{\rm yr^{-1}~km~s^{-1}})$ = $10^{-4.24\pm0.32}$ $(L_{\rm 
bol}/L_\odot)^{0.67\pm0.13}$ (Maud et al. 2015, for a sample of massive 
protostars).  The empirical correlations with cm emission described by 
equations \ref{eq:lbol} and \ref{eq:dotp} result in an expected 
correlation $(\dot P/M_{\sun}~{\rm yr^{-1}~km~s^{-1}})$ = 
$10^{-4.77\pm0.14}$ $(L_{\rm bol}/L_\odot)^{0.58\pm0.08}$, which is in 
agreement with the results obtained directly from the observations. We 
interpreted the correlation of the outflow momentum rate with the radio 
luminosity (eq. \ref{eq:dotp}) as a consequence of the shock ionization 
mechanism working in radio jets, and the correlation of the bolometric 
and radio luminosities (eq. \ref{eq:lbol}) as a consequence of the 
accretion and outflow relationship. In this context, the well-known 
correlation between the momentum rate of molecular outflows and the 
bolometric luminosity of their driving sources can be interpreted as a 
natural consequence of the other two correlations.

\section{Additional topics}

\subsection{Fossil outflows}

In the case of regions of massive star formation there are also clear 
examples of jets that show the morphology and spectral index 
characteristic of this type of sources. Furthermore, these massive jets 
fall in the correlations previously discussed.

However, there is a significant number of molecular outflows in regions 
of massive star formation where it has not been possible to detect the 
jet. Instead, ultracompact HII regions are found near the center of 
these outflows (e.g., G5.89-0.39, Zijlstra et al. 1990; G25.65+1.05 and 
G240.31+0.07, Shepherd \& Churchwell 1996; G45.12+0.13 and G45.07+0.13, 
Hunter et al. 1997; G192.16$-$3.82, Devine et al. 1999; 
G213.880$-$11.837, Qin et al. 2008; G10.6-0.4, Liu et al. 2010; 
G24.78+0.08, Codella et al. 2013; G35.58$-$0.03, Zhang et al. 2014).

We can think of two explanations for this result. One is that the 
central source has evolved and the jet has been replaced by an 
ultracompact HII region. From momentum conservation the outflow will 
continue coasting for a large period of time, becoming a fossil outflow 
in the sense that it now lacks an exciting source of energy. The 
alternative explanation is that a centimeter jet is present in the 
region but that the much brighter HII region makes it difficult to 
detect it. This is a problem that requires further research.

It should also be noted that in two of the best studied cases, 
G25.65+1.05 and G240.31+0.07, high angular resolution radio observations 
(Kurtz et al. 1994; Chen et al. 2007; Trinidad et al. 2011) have 
revealed fainter sources in the region that could be the true energizing 
sources of the molecular outflows. If this is the case, the outflows 
cannot be considered as fossil since they would have an associated 
active jet.

\subsection{Jets or ionized disks?}

\begin{figure}[h!]
\begin{center}
\vspace{0.4cm}
\includegraphics[width=0.48\textwidth]{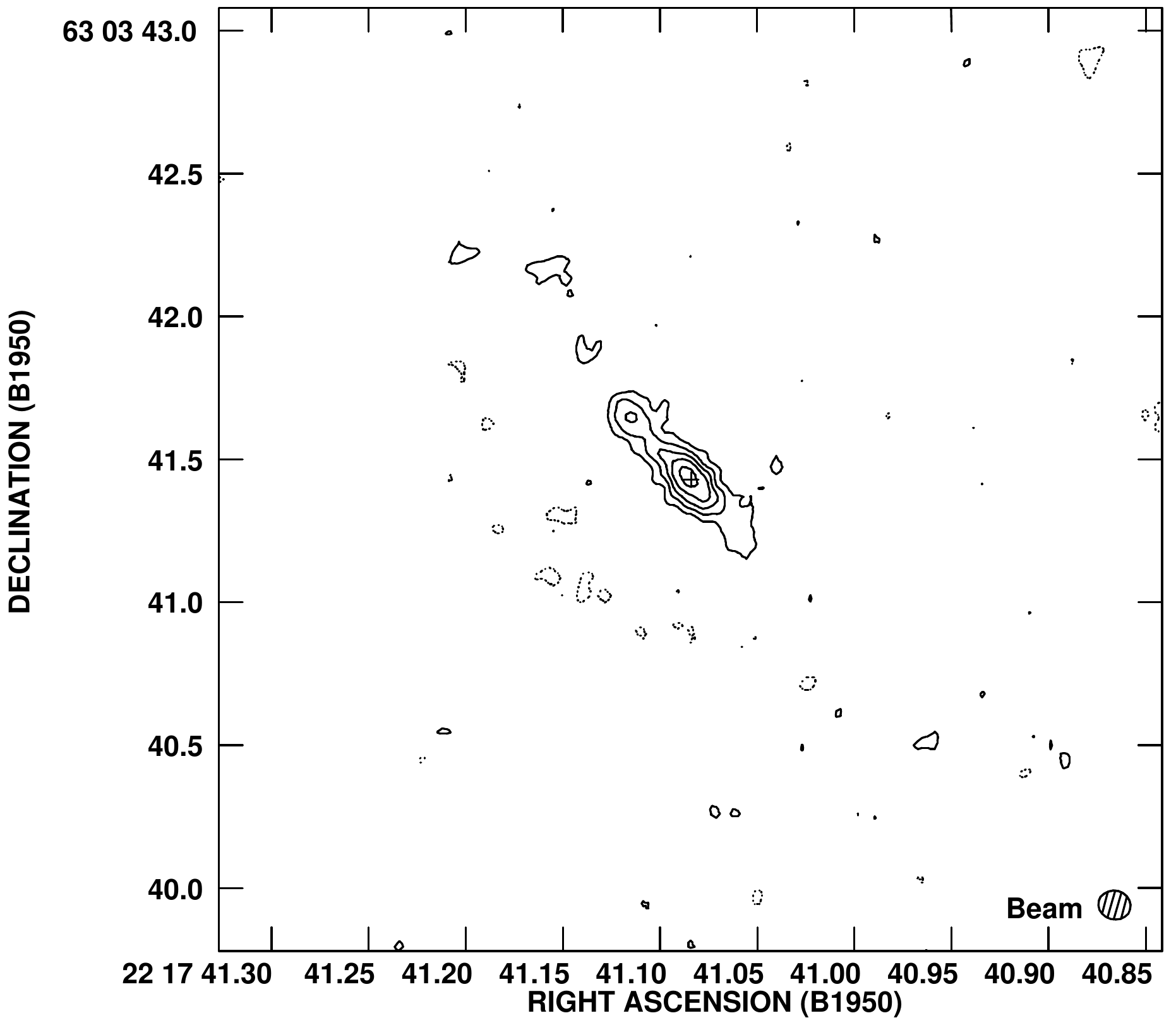} 
\hfill
\includegraphics[width=0.49\textwidth]{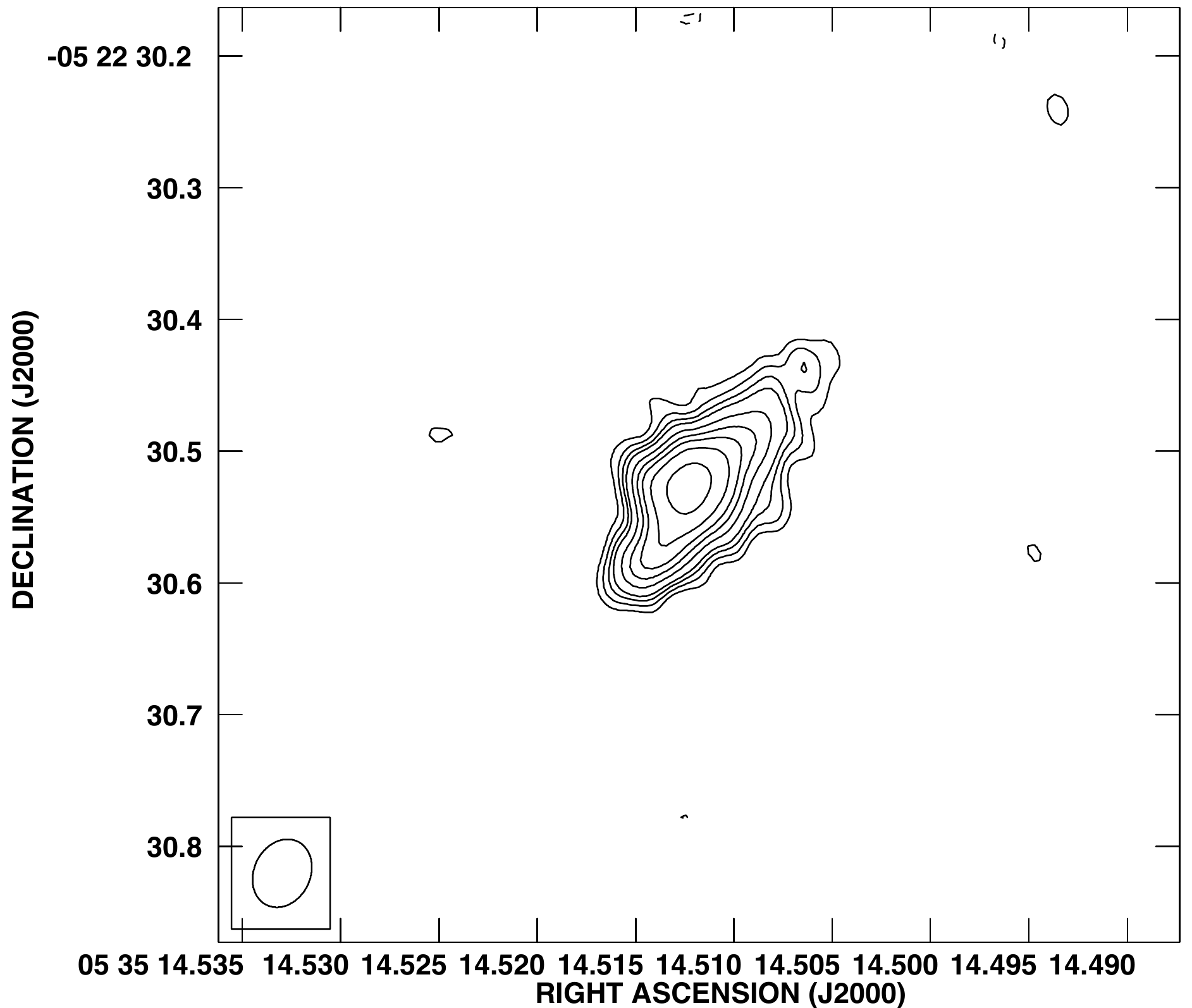}
 \caption{The radio sources associated with S140 IRS1 (left) and with 
source I in Orion (right). These sources of radio continuum emission are 
believed to trace photoionized disks and not jets. Image reproduced with 
permission from [left] Hoare (2006), copyright by AAS; and [right] based 
on Reid et al. (2007).
 \label{fig:iondisk} 
}
\end{center}
\end{figure}

The presence in star forming regions of an elongated centimeter source 
is usually interpreted as indicating the presence of a thermal jet. This 
interpretation is typically confirmed by showing that the outflow traced 
at larger scales by molecular outflows and/or optical/IR HH objects 
aligns with the small scale radio jet. In the sources where the true 
dust disk is detected and resolved, it is found to align perpendicular 
to the outflow axis. However, in a few massive objects there is evidence 
that the elongated centimeter source actually traces a photoionized disk 
(S106IR: Hoare et al. 1994; S140-IRS: Hoare 2006; Orion Source I: Reid 
et al. 2007; see Fig.~\ref{fig:iondisk}).  These objects show a similar 
centimeter spectral index to that of jets and one cannot discriminate 
using this criterion. There is also the case of NGC 7538 IRS1, a source 
that has been interpreted as an ionized jet (Sandell et al. 2009) or 
modeled as a photoionized accretion disk (Lugo et al. 2004), although it 
is usually referred to as an ultracompact HII region (Zhu et al. 2013).

A possible way to favor one of the interpretations is to locate the 
object in a radio luminosity versus bolometric luminosity diagram as has 
been discussed above. It is expected that a photoionized disk will fall 
in between the regions of thermal jets and UC HII regions in such a 
diagram. Another test to discriminate between thermal jets and 
photoionized disks would be to eventually detect radio recombination 
lines from the source. In the case of photoionized disks lines with 
widths of tens of km s$^{-1}$ are expected, while in the case of thermal 
jets the lines could exhibit widths an order of magnitude larger. It 
should be noted, however, that in the case of very collimated jets the 
velocity dispersion and, thus, the observed line widths, will be narrow.

\begin{figure}[h!]
\begin{center}
\includegraphics[width=0.9\textwidth]{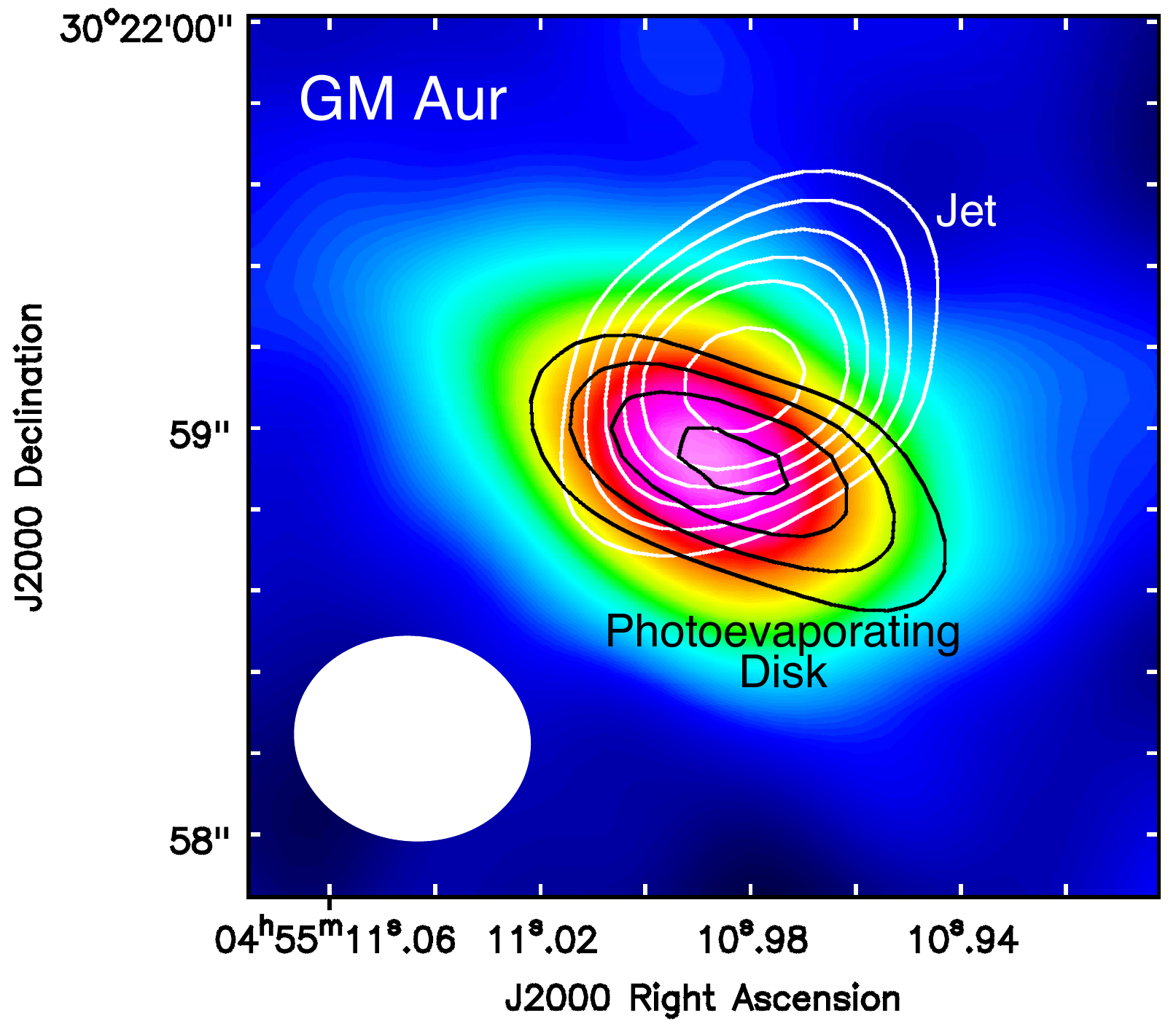} 
 \caption{Decomposition of the emission of GM Aur at cm wavelengths. The 
free-free emission at 3 cm of the radio jet is shown in white contours 
and that of the photoevaporative wind from the disk is shown in black 
contours. The dust emission from the disk at 7 mm is shown in color 
scale. The free-free emission of the two ionized components was 
separated by fitting two Gaussians to the 3 cm image after subtraction 
of the contribution at 3 cm of the dust of the disk, estimated by 
scaling the 7 mm image with the dust spectral index obtained from a fit 
to the spectral energy distribution. Image reproduced with permission 
from Mac\'\i as et al.\ (2016), copyright by AAS.
 \label{fig:gmaur} 
}
\end{center}
\end{figure}

The possibility of an ionized disk is also present in the case of low 
mass protostars. The high resolution images of GM Aur presented by 
Mac\'\i as et al. (2016) show that, after subtracting the expected dust 
emission from the disk, the centimeter emission from this source is 
composed of an ionized radio jet and a photoevaporative wind arising 
from the disk perpendicular to the jet (see Fig.~\ref{fig:gmaur}). It is 
believed that extreme-UV (EUV) radiation from the star is the main 
ionizing mechanism of the disk surface. Dust emission at cm wavelengths 
is supposed to arise mainly from grains that have grown up to reach 
pebble sizes (e.g., Guidi et al. 2016).

\section{Conclusions}

The study of jets associated with young stars has contributed in an 
important manner to our understanding of the process of star formation. 
We list below the main conclusions that arise from these studies.

1. Free-free radio jets are typically found in association with the 
forming stars that can also power optical or molecular large-scale 
outflows. While the jets trace the outflow over the last few years, the 
optical and molecular outflows integrate in time over centuries or even 
millenia.

2. The radio jets provide a means to determine accurately the position 
and proper motions of the stellar system in regions of extremely high 
obscuration.

3. The core of these jets emits as a partially optically thick free-free 
source. However, in knots along the jet (notably in the more massive 
protostars) optically thin synchrotron emission could be present. 
Studies of this non-thermal emission will provide important information 
on the role of magnetic fields in these jets.

4. At present there are only tentative detections of radio recombination 
lines from the jets. Future instruments such as SKA and the ngVLA will 
allow a new avenue of research using this observational tool.

5. The radio luminosity of the jets is well correlated both with the 
bolometric luminosity and the outflow momentum rate of the optical or 
molecular outflow. This is a result that can be understood theoretically 
for sources that derive most of its luminosity from accretion and where 
the ionization of the jet is due to shocks with the ambient medium. 
These correlations extend from massive young stars to the sub-stellar 
domain, suggesting a common formation mechanism for all stars.

\begin{acknowledgements}

GA acknowledges support from MINECO (Spain) AYA2014-57369-C3-3-P and 
AYA2017-84390-C2-1-R grants (co-funded by FEDER). LFR acknowledges 
support from CONACyT, Mexico and DGAPA, UNAM. CC-G acknowledges support 
from UNAM-DGAPA-PAPIIT grant numbers IA102816 and IN108218. We thank an 
anonymous referee for his/her useful comments and suggestions.

\end{acknowledgements}

\end{document}